\documentclass{svmult}
\usepackage{makeidx}         % allows index generation
\usepackage{graphicx}        % standard LaTeX graphics tool
\usepackage{multicol}        % used for the two-column index
\usepackage[bottom]{footmisc}% places footnotes at page bottom
\makeindex             % used for the subject index
\usepackage{epsfig}
\usepackage{graphics}% Include figure files
\usepackage{dcolumn}% Align table columns on decimal point
\usepackage{bm}% bold math

\def\cm-1{cm$^{-1}$}
\def\sco{Sr$_{14}$Cu$_{24}$O$_{41}$\,}
\def\scco{Sr$_{14-x}$Ca$_{x}$Cu$_{24}$O$_{41}$\,}
\def\lcco{La$_{6}$Ca$_{8}$Cu$_{24}$O$_{41}$\,}
\def\slcco{(Sr,La)$_{14-x}$Ca$_{x}$Cu$_{24}$O$_{41}$\,}
\def\sccoeight{Sr$_{6}$Ca$_{8}$Cu$_{24}$O$_{41}$\,}

\begin{document}

\title*{Collective Spin and Charge Excitations in  \slcco Quantum Spin Ladders}
\author{A. Gozar\inst{1,2} \and G. Blumberg\inst{1,*}}
\institute{
Bell Laboratories, Lucent Technologies, Murray Hill, NJ 07974, USA \and 
University of Illinois at Urbana-Champaign, Urbana, IL 61801, USA
\texttt{
\begin{center}
    (\textit{Frontiers in Magnetic Materials} (Ed. A.V. Narlikar), 
    Spinger-Verlag 2005, pp. 653-695).
\end{center}
}} 
\maketitle

\section{\slcco: the Structure and General Properties}
\label{sec:1}

In 1988, material research focussed around the study of high temperature superconducting copper-oxides brought about new phases of Cu-O based systems, the two-leg spin-ladders (2LL's), that have the general formula (A$_{1-x}$A'$_{x}$)$_{14}$Cu$_{24}$O$_{41}$, with A an alkaline earth metal and A' a trivalent (transition or lanthanoid) metal \cite{McCarronMRB88,SiegristMRB88}.
There were well-founded hopes that these materials could provide useful insight for the unresolved problems posed by the 2D cuprates \cite{DagottoScience96,DagottoRPP99} and from this perspective two main reasons triggered the interest of the scientific community.
One of them was based on a number of physical properties that are common for both ladders and high T$_{c}$'s.
These include the presence of similar Cu-O-Cu antiferromagnetic (AF) correlations which give rise to a finite spin gap and were predicted to generate $d$-wave like pairing of doped carriers \cite{DagottoPRB92RiceEL93}, the evidence for 'pseudogap' phenomena in optical absorption spectra \cite{OsafunePRL99} and, most importantly, the discovery of superconductivity under pressure evolving with hole doping in the AF environment \cite{UeharaJPSJ96,MaekawaNature96}.
The second reason resides in the crystal similarities and more precisely the fact that one can imagine building the 2D square Cu-O lattice by gradually increasing the coupling between individual 2LL's  \cite{SachdevScience00}, the simplicity of the latter making them more tractable for theoretical analysis.

\begin{figure}[t]
\centerline{
\epsfig{figure=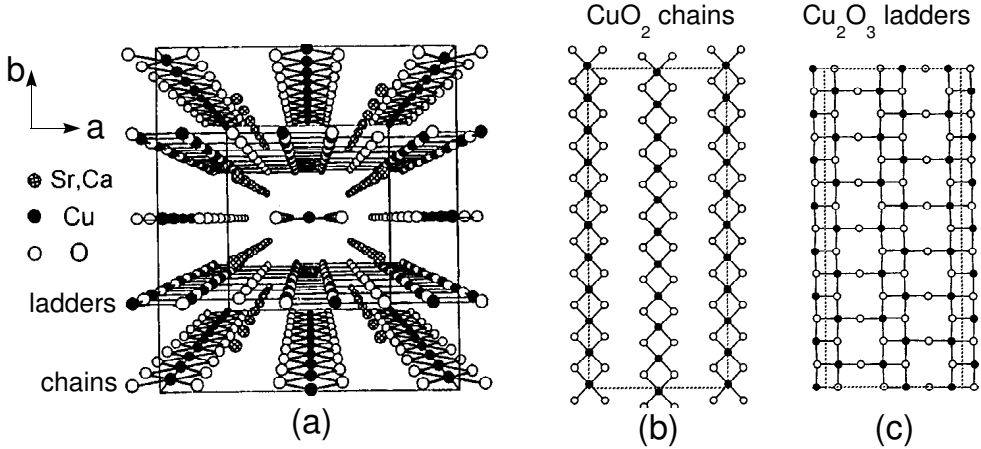,width=130mm}
}
\caption{
(a) The 3D structure of Sr$_{14}$Cu$_{24}$O$_{41}$ viewed in the $(ab)$ plane;
(b) the CuO$_{2}$ chains.
(c) the Cu$_{2}$O$_{3}$ two-leg ladders.
In (b) and (c) the black dots are Cu atoms and the empty circles represent O atoms.
}
\label{f11}
\end{figure}

The unit cell of Sr$_{14}$Cu$_{24}$O$_{41}$ contains four formula units, 316 atoms in all, this large number of atoms being due to the presence of two nearly commensurate substructures: the CuO$_{2}$ chains and the Cu$_{2}$O$_{3}$ 2LL's. 
A better understanding of the two interacting blocks can be achieved by decomposing the chemical formula into (Sr$_{2}$Cu$_{2}$O$_{3}$)$_{7}$(CuO$_{2}$)$_{10}$: planes of CuO$_{2}$ chains are stacked alternately with planes of Cu$_{2}$O$_{3}$ ladders and these are separated by Sr buffer layers, see Fig.~\ref{f11}.
The lattice constants of the individual sub-systems satisfy the approximate relation $7\ c_{ladder} \approx 10\ c_{chain}$.
The $b$-axis is perpendicular to the Cu-O layers which define the $(ac)$ plane, the $c$-axis being along the ladder/chain direction.
A valence counting shows that Sr$^{2+}_{14}$Cu$_{24}$O$^{2-}_{41}$ is intrinsically doped, the average valence per Cu atoms being $+ 2.25$.
Optical conductivity \cite{OsafunePRL97}, X-ray absorption \cite{NuckerPRB00}, $dc$ resistivity and magnetic susceptibility \cite{KatoPhysicaC96} measurements, as well as evaluations of the Madelung potential \cite{MizunoPhysicaC97} and valence-bond-sums \cite{KatoPhysicaC96} support the idea that in this compound the holes reside mainly in the chain structures and the isovalent Ca substitution for Sr in Sr$_{14-x}$Ca$_{x}$Cu$_{24}$O$_{41}$ induces a transfer of holes into the more conductive ladders.
A relatively large ladder carrier density change from 0.07 hole per Cu for $x = 0$ to about 0.2 for $x = 11$ due to Sr substitution was inferred from low energy optical spectral weight transfer\cite{OsafunePRL97}, but X-ray absorption \cite{NuckerPRB00}, while still supporting the hole migration scenario, is in favor of a less pronounced hole transfer.
On the other hand, La$^{3+}$ and Y$^{3+}$ substitutions for Sr decrease the total hole concentration, the \lcco compound containing no holes per formula unit.
As a result, the ladder systems provide the opportunity to study not only magnetism in low dimensional quantum systems like undoped ladders but also competing ground states and carrier dynamics in an antiferromagnetic environment.
Data interpretation, encumbered by the presence of two interacting subsystems in \slcco crystals, is being helped by experimental realizations of other related compounds like SrCu$_{2}$O$_{3}$, which contains only 2LL planes (Fig.~\ref{f11}c), or Sr$_{2}$CuO$_{3}$ and SrCuO$_{2}$, which incorporates only quasi-1D Cu-O chain units with a similar coordination as in Fig.~\ref{f11}b.
Unfortunately, doping in these latter systems has not been achieved so far.

Ca substitution in \scco has an important impact on the transport properties because of the chain-ladder hole transfer.
Indeed, while Sr$_{14}$Cu$_{24}$O$_{41}$ is an insulator showing an activation gap $\Delta \approx 2100$~K (180~meV), a crossover from insulating to metallic conduction at high temperatures takes place around $x = 11$ and for $x = 12$ the $c$-axis $dc$ resistivity has a minimum around 70~K separating quasi-linear metallic (above T~=~70~K) and highly insulating behavior at low temperatures \cite{OsafunePRL99}.
At higher Ca concentrations superconductivity under pressure has been observed, for example, a T$_{c}$ of 12~K under a pressure P~=~3~GPa was found in x~=~13.6 Sr$_{14-x}$Ca$_{x}$Cu$_{24}$O$_{41}$ \cite{UeharaJPSJ96}.

These properties, many of them common also to the 2D superconducting cuprates, underscore the potential value of the ladder systems for the understanding of superconductivity and also for the problem of identifying possibly competing order parameters in doped Mott-Hubbard systems.
The plan for this chapter is to present the magnetic properties of $S = 1/2$ 2LL's along with our Raman scattering data on the two-magnon (2M) excitation in Sr$_{14}$Cu$_{24}$O$_{41}$, showing its polarization, resonance and relaxation properties.
This is followed by the the analysis of Ca substitution effects on the low and high energy charge/spin degrees of freedom, our data supporting a scenario involving density-wave fluctuations as one of the competing orders for superconductivity.

%%%%%%%%%%%%%%%%%%%%%%%%%%%%%%%%%%%%%%%%%%%%%%%%%%%%%%%%%%%
\section{Magnetic Properties of Sr$_{14}$Cu$_{24}$O$_{41}$}
%%%%%%%%%%%%%%%%%%%%%%%%%%%%%%%%%%%%%%%%%%%%%%%%%%%%%%%%%%%

%%%%%%%%%%%%%%%%%%%%%%%%%%
\subsection{Energy Scales}
%%%%%%%%%%%%%%%%%%%%%%%%%%

Responsible for the magnetic properties are the Cu atoms which carry a spin $S = 1/2$ due to a missing electron on their $3d$ shells.
The AF super-exchange between them is mediated by the O ligand $2p$ orbitals.
The optical absorption due to transitions across the charge-transfer gap (determined by the energy difference between the Cu$3d$ and O$2p$ orbitals) is seen to occur at around 2~eV \cite{OsafunePRL97}.
The sign of the super-exchange as a function of the Cu-O-Cu bond angle can be qualitatively estimated semi-empirically as the balance of two terms: the first term is a relatively small, weakly bond angle dependent, ferromagnetic interaction while the second is antiferromagnetic, large for a 180$^{\circ}$ Cu-O-Cu bond but strongly varying with the bond angle, tending to zero around 90$^{\circ}$ \cite{AndersonRSV1C2}.

{\bf Cu-O chains --} As a result of nuclear magnetic/quadrupole resonance (NMR/NQR) \cite{TakigawaPRB98}, X-ray \cite{FukudaPRB02} and inelastic neutron scattering (INS) \cite{EcclestonPRL98,RegnaultPRB99} measurements,  the following picture provide clarification over some controversial aspects regarding charge/spin ordering in these structures.
NMR/NQR data identified two Cu$_{chain}$ sites, one carrying spin $1/2$ and one non-magnetic because of Zhang-Rice (ZR) singlet formation, that is a spin $S = 0$ state made out of a O$2p$ hole and a Cu$3d$ hole due to orbital hybridization. The data suggested the existence of a superstructure from the multipeak structure of the NMR spectra below about 150~K \cite{TakigawaPRB98}.
\begin{figure}[t]
\centerline{
\epsfig{figure=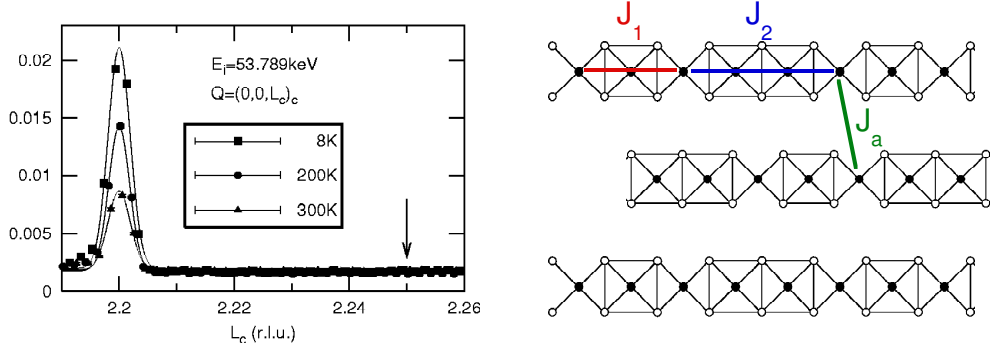,width=130mm}
}
\caption{
(a) X-ray intensity of the super structure peak seen at (0,0,2.2) in a $\theta-2\theta$ scan for several temperatures indicating a 5-fold charge modulation, from Ref.~\cite{FukudaPRB02}.
Q is the momentum transfer measured in chain reciprocal units.
(b) Charge pattern in the chains and the associated AF interactions as determined from Ref.~\cite{RegnaultPRB99}.
The squares denote Zhang-Rice singlets.
}
\label{f12}
\end{figure}
X-ray studies \cite{FukudaPRB02} established a five-fold charge modulation in the chains' ground state along the $c$ direction which exists at all temperatures below 300~K and a correlation length longer than 200~\AA, confirming an ordered pattern involving AF spin dimers separated by two ZR singlets, see Fig.~\ref{f12}.
Neutron scattering further supports such a superstructure by analyzing magnetic excitations out of the chain structures and evaluates the dominant intra-dimer exchange to be J$_{1} \approx 10$~meV \cite{EcclestonPRL98,RegnaultPRB99} which is also sets the value of spin gap in the dimerized chain.
Surprisingly, the inter-dimer and inter-chain exchanges were found to be of the same order of magnitude but of different signs: J$_{2} \approx$~-1.1~meV and J$_{a} \approx$~=~1.7~meV \cite{RegnaultPRB99} and, consistent with NMR/NQR data \cite{TakigawaPRB98}, 2D spin correlations due to J$_{a}$ were shown to develop below a characteristic temperature of about 150~K.  
Notable is the fact that if the ZR complexes are effectively made of truly Cu$^{3+}$ ions, the modulation shown in Fig.~1.2 would correspond to a Cu$_{chain}$ valence of 2.6+ meaning that all the holes are located in the chains.
Residual carriers are however present in the ladders and microwave \cite{KitanoEL01} and NMR/NQR \cite{TakigawaPRB98} data suggested the possibility of charge ordering in these systems too.

\begin{figure}[b]
\centerline{
\epsfig{figure=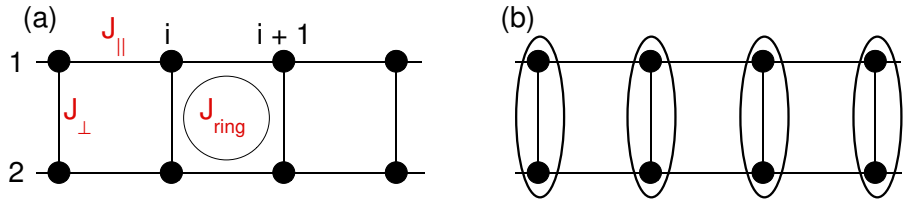,width=120mm}
}
\caption{
(a) Schematic of a two-leg ladder.
J$_{\parallel}$ and J$_{\perp}$ represent the nearest neighbor AF exchanges along the ladder legs and rungs respectively.
The circle is an example of a higher order spin interaction, in this case the ring exchange J$_{ring}$, which is thought to play an important role in the magnetic dynamics of the ladders, see text.
(b) The ground state of the two-leg ladder in (a) in the limit $y = J_{\parallel}/J_{\perp} \rightarrow 0$.
The ovals represent spin singlet states: $1/\sqrt{2} (|\uparrow \downarrow> - |\downarrow \uparrow>)$
}
\label{f13}
\end{figure}

{\bf Cu-O ladders --} At low temperatures Sr$_{14}$Cu$_{24}$O$_{41}$ can be regarded as an example of a 2LL structure close to half-filling (undoped with carriers).
Moreover, an individual 2LL, shown in Fig.~\ref{f13} is expected to incorporate the essence of the spin dynamics in this subsystem.
This is because the Cu-O-Cu bonds which are close to 180$^{\circ}$ generate a strong super-exchange J$_{\parallel}$ and J$_{\perp}$ (see Fig.~\ref{f13}) of the order of 130~meV ($\approx$~1000\cm-1).
This value is about two orders of magnitude stronger than the (frustrated) ferromagnetic inter-ladder interaction, see Fig.~\ref{f11}.
From the 2D cuprates experience, an expected Raman signature at energies of several J's is a two magnon (2M) like excitation consisting of a pair of spin-flips.
Low temperature behavior seen in magnetic susceptibility and NMR data show that, unlike in the cuprates, the low frequency spin behavior is not determined by gapless spin-wave modes, expected when one ignores small anisotropies which can create long wavelength gaps. 
Here there is a substantial spin-gap from the singlet ground state to the lowest triplet ($S = 1$) excitation.
The gap value for Sr$_{14}$Cu$_{24}$O$_{41}$ extracted from the temperature dependent Knight shift in Cu-NMR data was $\Delta_{S} \approx 32$~meV (260~\cm-1) \cite{TakigawaPRB98,MagishiPRB98}, in good agreement with the gap extracted from neutron scattering data \cite{EcclestonPRL98} in the same material as well as with the quasi-activated magnetization data [$\chi(T) \propto (1/\sqrt{T}) e^{-\Delta/k_{B}T}$ see Ref.~\cite{DagottoScience96}] in the 2LL SrCu$_{2}$O$_{3}$ \cite{AzumaPRL94}.
Spin-gap determination from magnetization measurements in Sr$_{14}$Cu$_{24}$O$_{41}$ is more ambiguous due to the prominent contribution from the chains.  
The magnetic properties of the Sr$_{14}$Cu$_{24}$O$_{41}$ ladders, is the concern of the following sections.

\subsection{Undoped Two-Leg Ladders: Theoretical Aspects}

\begin{figure}[b]
\centerline{
\epsfig{figure=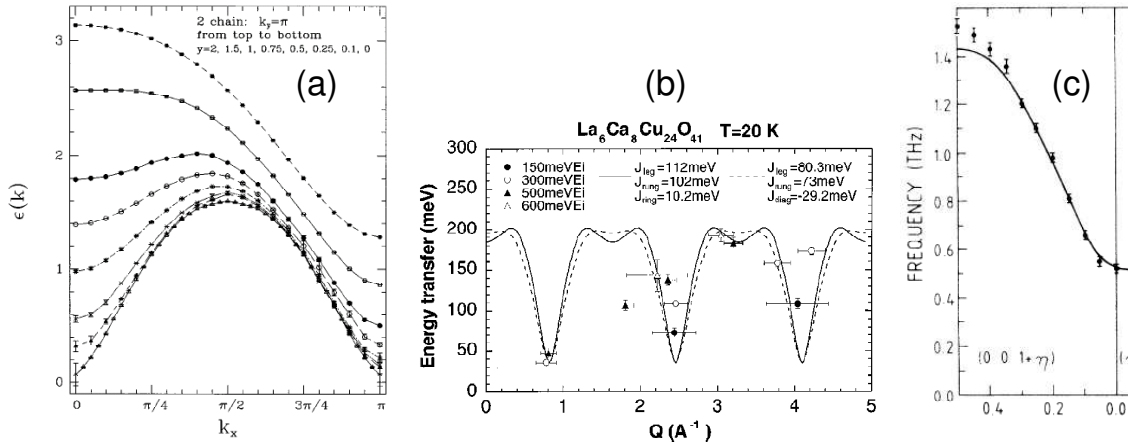,width=150mm}
}
\caption{
(a) Results of series expansions calculations (Ref.~\cite{OitmaaPRB96}) around the Ising limit for a two-leg ladder Hamiltonian, Eq.(1.1).
From top to bottom the elementary one-magnon excitation was computed for a decreasing ratio $J_{\perp}/J_{\parallel}$ from 2 to 0.
The limit $J_{\perp}/J_{\parallel} \rightarrow 0$ corresponds to spinons, the limit $J_{\perp}/J_{\parallel} \rightarrow \infty$ corresponds to uncoupled dimers on the ladder rungs.
(b) Neutron scattering results for the elementary triplet dispersion in \lcco (from Ref.~\cite{MatsudaJAP00})
(c) Dispersion inferred from neutron scattering data (Ref.~\cite{BuyersPRL86}) along the chain direction in the Haldane system CsNiCl$_{3}$.
The similarity with Fig.~4a becomes obvious in the strong coupling limit, J$_{\perp} >$~J$_{\parallel}$. 
}
\label{f14}
\end{figure}
The starting point for the determination of the ladder excitation spectrum has been the AF nearest-neighbor isotropic Heisenberg Hamiltonian, allowing for the leg and rung couplings J$_{\parallel}$ and J$_{\perp}$, see Fig.~\ref{f13}a.
This Hamiltonian reads:
\begin{equation}
H = H_{\parallel} + H_{\perp} = J_{\parallel} \sum_{i,\alpha = 1,2} {\bf S}_{i,\alpha} \cdot {\bf S}_{i+1,\alpha} + J_{\perp} \sum_{i} {\bf S}_{i,1} \cdot {\bf S}_{i,2} 
\label{e11}
\end{equation}
From the crystal structure one can anticipate that the relevant parameter range for the leg to rung super-exchange ratio is $ y = J_{\parallel}/J_{\perp} \approx 1$.
The excitation spectrum could be easily understood starting from the strong coupling limit, $J_{\parallel}/J_{\perp} \rightarrow 0$: the ground state is a simple product of singlets sitting on each rung, see Fig~\ref{f13}b.
Excited N-particle states (where N is the number of triplets) are highly degenerate and are obtained by exciting elementary triplets on N different rungs \cite{DagottoScience96,DagottoRPP99,DagottoPRB92RiceEL93,BarnesPRB93}.
The nature of the ground and first excited states evolves smoothly when a small J$_{\parallel}$ is present.
This allows the rung triplets to propagate along the ladder giving rise to dispersion in the reciprocal space.
The bandwidth is proportional to J$_{\parallel}$ and the band minimum of the one-triplet branch is at the Brillouin zone boundary, $k = \pi$ \cite{BarnesPRB93}.
In the limit of uncoupled AF $S =1/2$ chains, $J_{\parallel}/J_{\perp} \rightarrow \infty$, the result is also known and the ground state is characterized by an algebraic decay of magnetic correlations, the excitation spectrum is gapless with soliton-like $S = 1/2$ excitations (spinons) \cite{FadeevPL81}.

The picture described above is supported by theoretical calculations, and it turns out that in the general case the 'physics' of undoped 2LL's is dominated by the strong coupling limit.
\begin{figure}[t]
\centerline{
\epsfig{figure=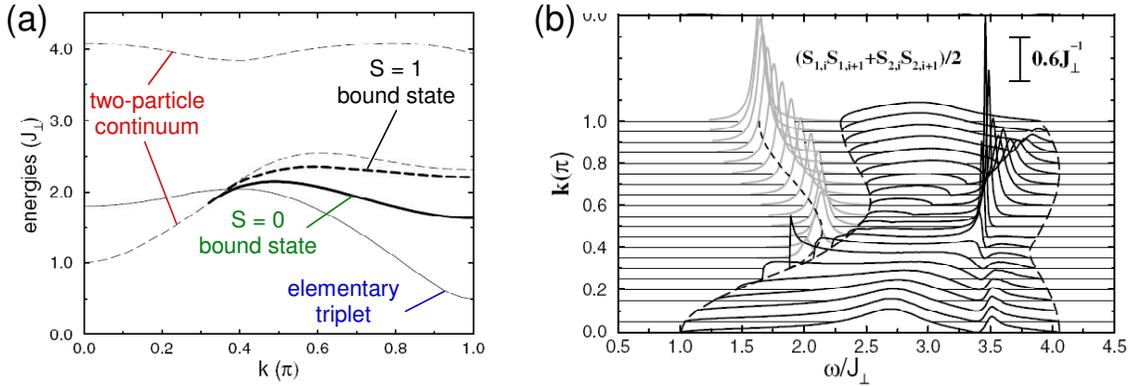,width=150mm}
}
\caption{
(a) Excitation spectrum in the one and two-particle channels at isotropic coupling $J_{\parallel}/J_{\perp} = 1$ from Ref.~\cite{KnetterPRL01}.
The elementary triplet, the two-particle continuum as well as the two-magnon singlet and triplet bound states are indicated in the figure.
(b) The $k$-resolved spectral density $I(k,\omega)$ of the ${\bf S}_{i,1} \cdot {\bf S}_{i+1,1} + {\bf S}_{i,2} \cdot {\bf S}_{i+1,2}$ operator.
This is proportional to the magnetic Raman response in parallel polarization with the electric field along the leg direction.
The divergences observed around $k \approx 0.6$ are due to the hump-dip structure of the $S = 0$ bound state in (a).
Data from Ref.~\cite{KnetterPRL01}.
}
\label{f15}
\end{figure}
\begin{itemize}
\item \emph{The ground state} is disordered and has exponential falloff of the spin-spin correlations.
A good description of the magnetic correlations is achieved within the resonance valence bond (RVB) model \cite{WhitePRL94}. 
For a pictorial representation see Fig.~\ref{f16}b.
Although a high $J_{\parallel}/J_{\perp}$ increases singlet correlations beyond nearest-neighbors, a ground state built up as a superposition of short-ranged resonating valence bonds remains a good approximation.
For odd-leg ladders long-ranged singlets must be included in the ground state description \cite{WhitePRL94}.
\item \emph{The one particle excitations} of the ladder have a gap $\Delta_{S}$ because any finite J$_{\perp}$ confines the $S = 1/2$ spinons binding them to an integer $S = 1$ 'magnon'.
Results of series expansions around the Ising limit for 2LL's at various couplings $y = J_{\parallel}/J_{\perp}$ from Ref.~\cite{OitmaaPRB96} are shown in Fig.~1.4a.
These results are further confirmed by exact diagonalizations \cite{DagottoPRB92RiceEL93}, numerical \cite{BarnesPRB93} and perturbative \cite{KnetterPRL01} analysis.
It has been also found that the spin gap remains finite for even leg ladders (although the gap decreases with increasing the number of legs) while odd-legged ladders are gapless and have a power law fall-off of spin-spin correlations \cite{WhitePRL94}.
This resembles the gapless and gapped alternance of the spectrum for isotropic AF half-integer and integer spin chains \cite{Haldane83}. 
The similarity is not accidental since a spin $S$ chain can be described as  $2S$ coupled spin $S = 1/2$ chains with appropriately chosen interchain coupling.
This analogy is beautifully confirmed by the dispersion found above the N\'{e}el temperature in an experimental realization of a Haldane system, CsNiCl$_{3}$, a quasi-1D nearly isotropic $S = 1$ AF chain \cite{BuyersPRL86}.
In Fig.~\ref{f14} we show for comparison the experimental elementary magnon dispersion in CsNiCl$_{3}$ along with experimental data and theoretical predictions for 2LL.
\item \emph{The two-particle states}: The elementary magnon branch will generate a two-magnon continuum starting from $2\Delta_{S}$ at $k = 0$.
In addition, this spectrum contains additional magnetic bound/antibound states.
These are states with discrete energies which are found below/above the two particle continuum \cite{TrebstPRL00andThesis}.
Bound states have been found in the singlet ($S = 0$), triplet ($S = 1$) and quintuplet ($S = 2$) sectors.
A typical excitation spectrum calculated perturbatively for isotropic coupling, $J_{\parallel}/J_{\perp} = 1$, and containing several types of two-particle excitations discussed above is shown in Fig.~\ref{f15}.
A particularity of 2LL's is the fact that the bound states 'peel off' the continuum at finite values of $k$.
The importance of higher order spin terms will be stressed in the following sections in connection with data analysis.
This analysis will show that one has to go beyond the nearest neighbor Heisenberg Hamiltonian of Eq.~(\ref{e11}) in order to explain the experimental data.
Regarding the question whether the best description at all energies is in terms of fractional or integer spin excitations, it is worth noticing that, at least in the limit $J_{\parallel}/J_{\perp} \leq 1$, there is no necessity to resort to fractional spin states.
A description in terms of truly bosonic excitations works well in the sense that spectral densities of spin-ladders can be described well by using integer spin excitations \cite{KnetterPRL01}.  
\end{itemize}

%%%%%%%%%%%%%%%%%%%%%%%%%%%%%%%%%%%%%%%%%%%%%%%%%%%%%%%%%%%%%%%%%%%%%%
\subsection{Low Temperature Two-Magnon Light Scattering in Sr$_{14}$Cu$_{24}$O$_{41}$}
%%%%%%%%%%%%%%%%%%%%%%%%%%%%%%%%%%%%%%%%%%%%%%%%%%%%%%%%%%%%%%%%%%%%%%

\begin{figure}[t]
\centerline{
\epsfig{figure=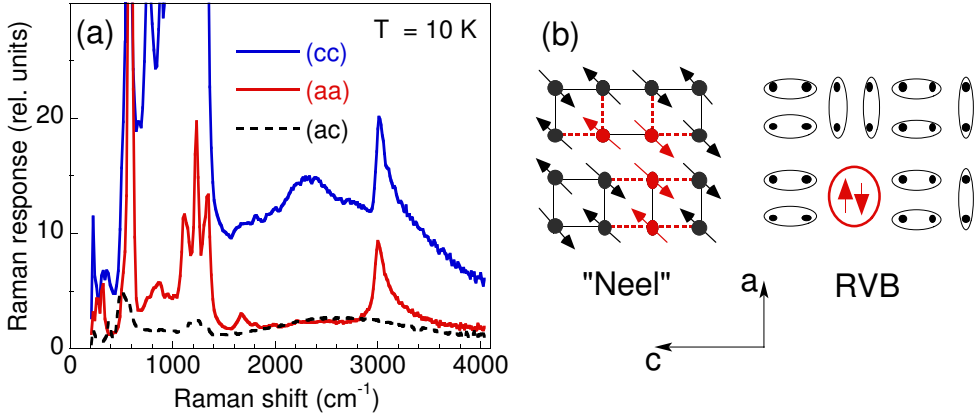,width=130mm}
}
\caption{
(a) Low temperature Raman spectra in several polarizations.
The sharp, asymmetric peak at 3000~\cm-1 is the two-magnon feature.
The strong features below about 1200~\cm-1 as well as the bump around 2350~\cm-1 (which has about four times the energy of the 580~\cm-1 Oxygen mode) in $(cc)$ polarization represent single and multi phonon excitations.
(b) Cartoon showing the two-magnon excitation.
Left: assuming a local AF N\'{e}el order the spin exchange can take place along the rungs/legs of the ladder.
Broken bonds and exchanged spins are shown in red.  
Right: a snapshot of the short-range RVB ground state which is a superposition of states like the top figure. 
The bottom picture represents a locally excited singlet state of two triplets.
}
\label{f16}
\end{figure}

In this section we will discuss symmetry, spectral and resonance properties of the 2M excitation in \sco at T~=~10~K.
Figure~\ref{f16} shows Raman spectra in $(cc)$, $(aa)$ and $(ac)$ polarization taken with an excitation energy $\omega_{in} =$~1.84~eV.
The spectra consist of a lower energy part where phonons are observed (see caption of Fig.~\ref{f16}) and a sharp asymmetric peak at 3000~\cm-1 present in parallel polarizations.
In both $(aa)$ and $(cc)$ polarizations the 3000~\cm-1 peak is situated at exactly the same energy.
In $(ac)$ polarization this feature is not present.
The energy of the 3000~\cm-1 mode, much larger than the relevant magnetic interactions in the chain structures, allows an unambiguous assignment of this excitation to the ladder systems.
A comparison with the 2D tetragonal cuprates \cite{GirshPRB96,SugaiPRB90} in terms of energy scales argues for the interpretation of the 3000~\cm-1 peak in terms of ladder 2M excitations.
Moreover, in 2D cuprates the 2M feature has B$_{1g}$ symmetry, this representation becoming the identical representation in the orthorhombic group to which the ladder structure belongs.
Indeed, as can be seen from Fig.~\ref{f16}, in \sco this excitation is fully symmetric.

Although for the 2D cuprates a semi-classical counting of broken magnetic bonds within a local N\'{e}el environment (see Fig.~\ref{f16}b) gives a good estimate ($3 J$) for the 2M energy (which is found by more elaborate calculations to be situated around $2.7\ J$), in 2LL's this approach is not suitable.
On one hand any small anisotropy in the exchange parameters $J_{\parallel}$ and $J_{\perp}$ should lead to different peak energies in $(aa)$ and $(cc)$ polarizations, see Fig.~\ref{f16}b, which is not observed, and on the other hand, even in the improbable case of less than 0.03\% anisotropy given by our energy resolution, this 'Ising counting' estimates $J \approx 200$~meV which is almost 50\% higher than the super-exchange in related 2D cuprates.
The failure of this approach may be related to the fact that the ground state of the 2LL's cannot be described classically. 
A RVB description of the ladder ground state has been proposed \cite{WhitePRL94}.
This can be understood as a coherent superposition of 'valence bonds', which are spin singlets, shown in Fig.~\ref{f16}c.
For even leg ladders the RVB states are short ranged (the singlets extend only over nearest neighbor Cu spins) and in this context, starting from an 'instantaneous configuration' of the ground state, the 2M excitation can be visualized as a state in which two neighboring singlets get excited into a higher energy singlet state made out of two triplet excitations.

\textbf{Symmetry --} The polarization selection rules for the 2M scattering can be explained using the effective spin Hamiltonian corresponding to the photon induced spin exchange process \cite{FleuryPR68,ShastryPRL90} which reads
\begin{equation}
H_{FL} \propto \sum_{<i,j>} ({\bf e}_{in} \cdot {\bf r}_{ij}) ({\bf e}_{out} \cdot {\bf r}_{ij}) {\bf S}_{i} \cdot {\bf S}_{j}
\label{e12}
\end{equation}
where  \textbf{S}$_{i}$, \textbf{S}$_{j}$ are Cu spins on the lattice sites $i$ and $j$, \textbf{r}$_{ij}$ in the vector connecting these sites and \textbf{e}$_{in}$/\textbf{e}$_{out}$ are the unit vectors corresponding to the incoming/outgoing polarizations.
The polarization prefactor shows that the 2M scattering should occur only in parallel polarizations, consistent with the experimental observations.

\textbf{Determination of J's --}
The problem of quantitatively estimating the magnitude of the super-exchange integrals is non-trivial in spite of the fact that there are several experimental techniques which probed magnetic excitations like neutron scattering \cite{EcclestonPRL98,MatsudaJAP00}, Raman \cite{SugaiPSS99,GozarPRL01} and IR spectroscopy \cite{WindtPRL01,NunnerPRB02}. 
For the latter technique, the authors claim that the strong mid-IR absorption features between 2500 and 4500~\cm-1 are due to phonon assisted 2M excitations.
The main problem was to reconcile by using only the Hamiltonian from Eq.~(\ref{e11}) the smallness of the zone boundary spin gap $\Delta_{S} = 32$~meV \cite{EcclestonPRL98} with respect to the magnitude of the one triplet energies close to the Brillouin zone center (see Ref.~\cite{MatsudaJAP00} and Fig.~\ref{f14}), which is thought to determine the position of the 2M Raman peak \cite{SchmidtPRL03} as well as the structure and the large energy range in which the mid-IR magnon absorption is seen \cite{WindtPRL01,NunnerPRB02}.
\begin{figure}[b]
\centerline{
\epsfig{figure=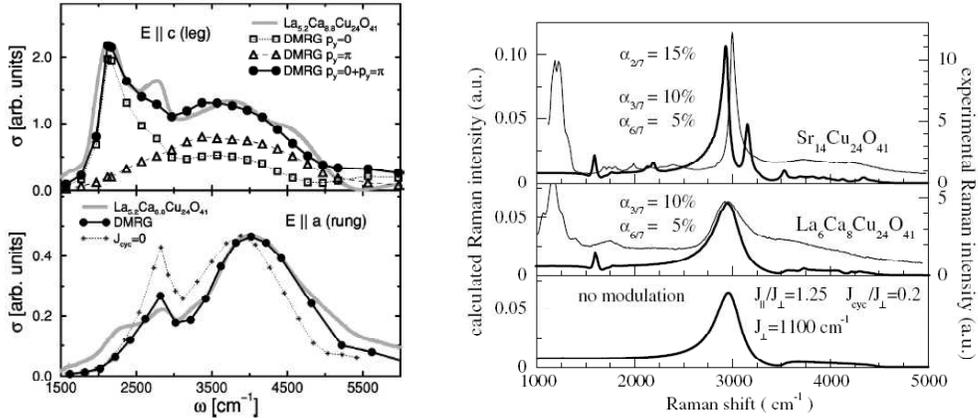,width=130mm}
}
\caption{
Left: IR absorption data (thick grey line) for two polarizations of the electric field along with theoretical calculations for the phonon assisted two-magnon absorption (from Ref. \cite{NunnerPRB02}).
The parameters used are: $y = J_{\parallel} / J_{\perp} = 1.3$, $J_{\perp} = 1000$ \cm-1 and a ring/cyclic exchange (see text and Fig.~\ref{f14}) $J_{ring} / J_{\perp} = 0.2$
Right: Calculated (thick lines) Raman response for the two-magnon scattering using a perturbative approach from Ref.~\cite{SchmidtPRL03}.
The parameters used are: $y = J_{\parallel} / J_{\perp} = 1.25$, $J_{\perp} = 1100$ \cm-1 and $J_{ring} / J_{\perp} = 0.2$.
The experimental data (thin lines) in \lcco and \sco are taken from Ref.~\cite{SugaiPSS99}.
The upper two panels are calculated using an additional super-modulation on the magnon dispersion due to the interaction induced by the charge ordering in the chain structures (see the text for discussion).
The parameter $\alpha$ specifies the kind of the superimposed supermodulation.
}
\label{f17}
\end{figure}
The proposed solution to this problem was to consider, besides $J_{\parallel}$ and $J_{\perp}$ the presence of a ring exchange $J_{ring}$ \cite{BrehmerPRB99}, which is a higher order spin correction whose effect can be understood as a cyclic exchange of the spins on a square plaquette determined by two adjacent ladder rungs, see Fig.~\ref{f14}.
The net effect of including such an interaction, which has the form $H_{ring} = 2 J_{ring} [({\bf S}_{1,i} \cdot {\bf S}_{1,i+1}) ({\bf S}_{2,i} \cdot {\bf S}_{2,i+1}) + ({\bf S}_{1,i} \cdot {\bf S}_{2,i}) ({\bf S}_{1,i+1} \cdot {\bf S}_{2,i+1}) - ({\bf S}_{1,i} \cdot {\bf S}_{2,i+1}) ({\bf S}_{1,i+1} \cdot {\bf S}_{2,i})]$, is to renormalize down the spin gap so that the ratio of the magnon energy at the zone boundary with respect to the one at the zone center is decreased.
The introduction of $J_{ring} \approx 0.1 J_{\perp}$ helped fitting the INS data (see Ref.~\cite{MatsudaJAP00} and Fig.~\ref{f14}) and an even higher ratio is able to better reproduce the experimental Raman and IR data (see Fig.~\ref{f17}).
The parameter sets used for the quantitative analysis of the spectroscopic data have $J_{\parallel} / J_{\perp}$ between 1.25 and 1.3 and a sizeable cyclic exchange, $J_{ring} / J_{\perp}$ of about 0.25~-~0.3.
The absolute value chosen for $J_{\perp}$ is 1000~-~1100~\cm-1.
Both the value of $J$ and $J_{ring}$ are quantitatively consistent with those inferred for the 2D AF cuprates~\cite{KataninPRB02}.
In the latter case, the cyclic exchange was used in order to reproduce the neutron scattering findings regarding the $k$ dependence of the energy of the one-magnon excitations in the proximity of the Brillouin zone boundary \cite{KataninPRB02}.
However, as opposed to the cuprates, the 2M seen in Fig.~\ref{f16} at 3000~\cm-1 cannot provide a direct determination of the super-exchange, even if no terms other than $J_{\parallel}$ and $J_{\perp}$ had to be included in the spin Hamiltonian.
This problem is related to the fact that, in spite of the theoretical results shown in Fig.~\ref{f17}b which suggest good agreement with the experiment, the spectral shape of the sharp 2M feature and its origin is still an open question; this issue will be discussed in the following.

\begin{figure}[b]
\centerline{
\epsfig{figure=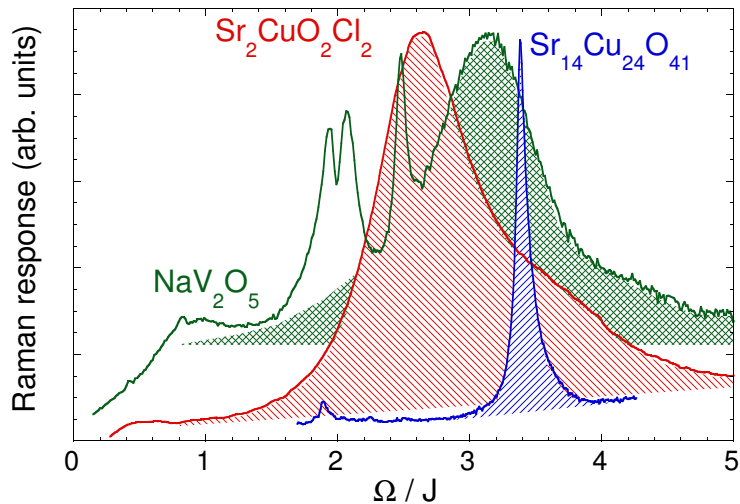,width=100mm}
}
\caption{
Magnetic Raman continua in several AF spin $S = 1/2$ systems.
Red: Sr$_{2}$CuO$_{2}$Cl$_{2}$ (a 2D square lattice with long range order).
Green: NaV$_{2}$O$_{5}$ (a two-leg ladder at quarter filling factor which can be mapped on a quasi 1D AF chain).
Blue: \sco, the excitation seen in Fig.~\ref{f17} but in this case taken with $\omega_{in} = 2.65$~eV incoming laser energy.
}
\label{f18}
\end{figure}

\textbf{Two-magnon relaxation --} 
While in the case of 2D cuprates theory has problems with explaining the large scattering width of the 2M excitation, in 2LL's the situation is reversed; this is one of the most interesting points made in Ref.~\cite{GozarPRL01}.
To emphasize the 2M sharpness, we compare it in Fig.~\ref{f18} to the corresponding excitation in Sr$_{2}$CuO$_{2}$Cl$_{2}$ which has one of the sharpest 2M feature among  2D AF copper oxides \cite{GirshPRB96} as well as to the multi-spinon scattering from a 2LL at quarter filling (which can be mapped on a quasi 1D $S = 1/2$ AF chain), as seen in the high temperature phase of NaV$_{2}$O$_{5}$.
For Sr$_{2}$CuO$_{2}$Cl$_{2}$ the FWHM is about 800~\cm-1 \cite{GirshPRB96} and this is comparable, in relative units, with the large scattering width observed for the spinon continuum.
In \sco the width is only about 90~\cm-1.
The 2M approximation for the magnetic light scattering in 2D cuprates, while giving a good estimate for the 2M peak energy, cannot reproduce its spectral profile.
This approximation makes the following three basic assumptions: 
\begin{itemize}
\item the ground state is a fully ordered N\'{e}el state; 
\item the spin pair excitations consist of states which have exactly two spins flipped with respect to the N\'{e}el configuration; 
\item since the light wavelength is much larger than the unit cell, only combinations of $(k,-k)$ magnons are allowed.
\end{itemize}
This approach neglects quantum fluctuations which means that the true ground state will also contain configurations of flipped spins and also that the spin-pair states will be admixtures of 2, 4, 6 ... spin flips in the ground state.
The narrow calculated width of the 2M was found, however, to be stable with respect to the inclusion of higher order spin interactions. 
Neither exact diagonalization nor Monte Carlo simulations were able to fully reproduce the 2M scattering width \cite{CanaliPRB92SandvikPRB98} although these calculations improved the results obtained within the 2M approximation.
It has been proposed by Singh \emph{et al.} in Ref.~\cite{SinghPRL89} that it is the quantum fluctuations effects inherent to the Heisenberg model with $S = 1/2$ which lead to the observed broadening.
The importance of intrinsic inhomogeneities and the role of phonons have also been invoked in the literature.

We were surprised that even in lower dimensionality (the structure determined by the 2LL's is quasi-1D), where the quantum fluctuations are expected to be stronger, the 2M Raman spectra display a narrow profile, a phenomenon which questions the importance attributed to these effects in low spin systems \cite{SinghPRL89}.
This prominent question triggered theoretical work, part of which is shown in Fig.~\ref{f17}.
The authors of Ref.~\cite{SchmidtPRL03} challenged our point and claimed a resolution in terms of both the existing quasi-commensuration between the unit cell constants of the chain and ladder structures ($7\ c_{ladder} \approx 10\ c_{chain}$) and the supermodulation induced by the charge order in the chain structures, which is shown in Fig.~\ref{f12}.
The calculation of the 2M Raman response without the modulation (lower panel in Fig.~\ref{f17}b) reveals indeed a broader 2M peak \cite{SchmidtEL01}, while inclusion of chain-ladder interaction renders a sharp 2M excitation because of the backfolding of the dispersion of the elementary triplet (Figs.~1 and 2 in Ref.~\cite{SchmidtPRL03}).
This opens gaps at the points of intersection with the supermodulation wavevectors and will have a drastic effects in the spectral shape because of the induced divergences in the density of states.

The agreement with the experimental data in Fig.~\ref{f17} is pretty good; however, these claims have recently been put to rest by a Raman experiment, Ref.~\cite{GoblingPRB03}, in the undoped 2LL compound SrCu$_{2}$O$_{3}$ (which contains no chains but only undoped 2LL's), experiment which revealed a 2M peak as sharp as in \sco.
This clearly shows that the sharpness is related neither to the interaction between the two substructures in \sco nor to the residual carriers in the 2LL structure of \sco but instead it is due to intrinsic 2LL's effects.
Two major differences between the 2LL's and 2D cuprates or 1D AF spin chains are the facts that in the former the low energy relaxation channels are suppressed due to the presence of a spin gap and also that the excitation spectrum of 2LL's supports the existence of magnetic bound states outside the continuum of excitations.
Although this may be a plausible explanation, the 2M singlet bound state peels off the continuum only at finite values of $k$, see Fig.~\ref{f15}, and besides that, the energy at $k = 0$ is too small ($2 \Delta_{S} = 64$~meV = 512~\cm-1) to account for the observed peak energy at 3000~\cm-1.
If the sharpness is from the hump-dip feature in the dispersion of the elementary triplet close to the Brillouin zone center, Fig.~\ref{f15}b and the corresponding Van Hove singularities, it seems that such divergences are found only at finite values of $k$ while at $k = 0$ the spectral density is quite broad \cite{KnetterPRL01}.
This is why we suggest here an explanation in relation to a possible spin density wave (SDW) modulation which is intrinsic to 2LL's and will lead to a backfolding of the magnon dispersion.
This effect is similar in spirit with the one proposed in Ref.~\cite{SchmidtPRL03} but this time due to intrinsic effects.
Regarding the asymmetry of the 2M feature it would also be worth considering multi-magnon interaction effects which may lead to the asymmetric Fano-like shape of the sharp 3000~\cm-1 feature due to the interaction with the underlying magnetic continuum.

Noteworthy is the resemblance of the elementary triplet dispersion in 2LL's and the $k$ dependence of the one-magnon excitation in La$_{2}$CuO$_{4}$ away from the Brillouin zone center.
There are several articles, some of them very recent \cite{TranquadaNature04}, which stress the failure of the spin wave models in 2D cuprates arguing that the 'physics' of magnetic excitations is fundamentally different at low and high energies: while semi-classical magnon theory holds at low energies, it has been argued that at short wavelengths the effect of fluctuations is more pronounced and the spin dynamics suggest an underlying structure similar to the one provided by 2LL's, which is due to a SDW-like modulation in the 2D planes.
Interestingly, the data in Sr$_{2}$CuO$_{2}$Cl$_{2}$ and NaV$_{2}$O$_{5}$ from Fig.~\ref{f18} suggest instead a more pronounced similarity to the magnetic scattering in 1D $S = 1/2$ AF chains.
It seems at this point that not only the 2M profile in 2LL's but also the one in 2D cuprates constitute open questions which have recently received renewed attention.
It would be very interesting if the physics in these two systems is found to be related to each other.

\begin{figure}[t]
\centerline{
\epsfig{figure=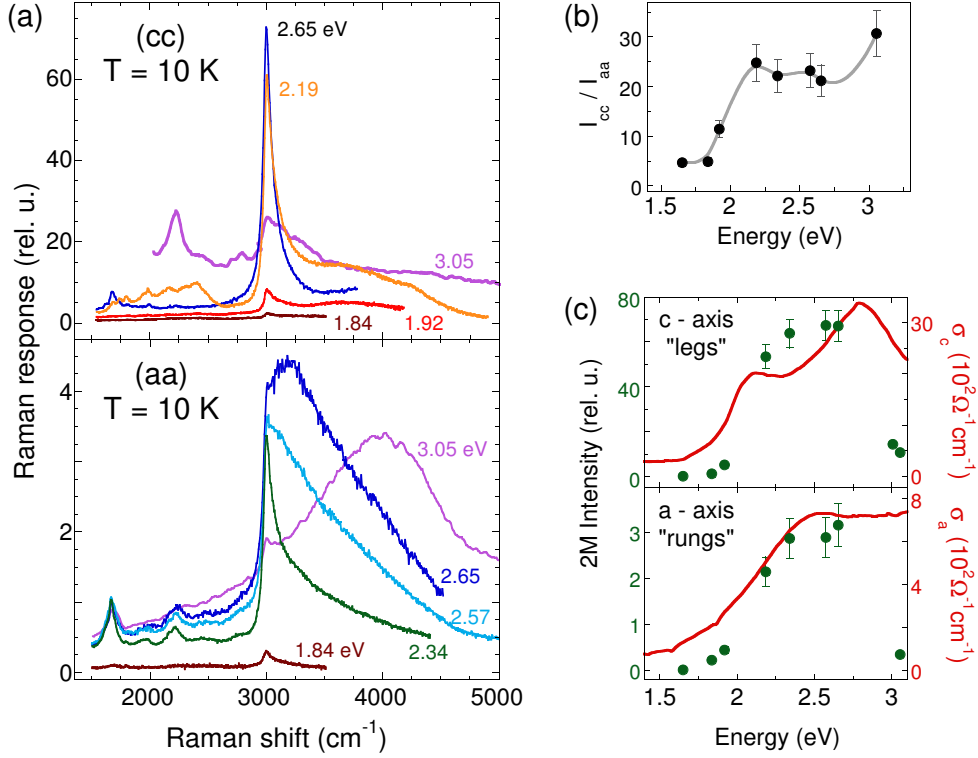,width=130mm}
}
\caption{
Two-magnon resonance profile in \sco at T = 10K.
(a) Data in $(cc)$ (upper panel) and $(aa)$ (lower panel) polarizations for different excitation energies $\omega_{in}$.
(b) The relative intensity of the 2M excitation in $(cc)$ compared to $(aa)$ polarization as a function of the incoming photon energy.
(c) The resonance profile of the 2M peak (filled circles, left scale) and the optical conductivity (solid line, right scale).
As in (a), the upper/lower panels refer to the data in which the electric field is parallel to the ladder legs/rungs. 
}
\label{f19}
\end{figure}

%%%%%%%%%%%%%%%%%%%%%%%%%%%%%%%%%%%%%%%%%
\textbf{Two-magnon excitation profile --}
%%%%%%%%%%%%%%%%%%%%%%%%%%%%%%%%%%%%%%%%%
A summary of our experimental study of the 2M dependence on the incoming photon energy is shown in Fig.~\ref{f19}.
Like the 2D cuprates, the Cu-O based ladders are known to be charge-transfer (CT)-type Mott insulators, the CT gap being determined by the energy difference between the Cu $3d$ and O $2p$ orbitals.
A Raman resonant study is interesting since, along with optical absorption, it gives information about the nature of the ground as well as of high energy electronic states across the CT gap.
This is because the photon induced spin exchange takes place in two steps: a photoexcited state consisting of an electron-hole pair is created by the interaction of the system in its ground state with an incoming photon and then this intermediate state collapses into an excited magnetic state characterized by broken AF bonds.
One expects therefore that such a process, in which the interaction with light occurs in the 2$^{nd}$ order perturbation theory, will show a strong dependence on the incoming photon energy \cite{GirshPRB96}.

This is what we observe in \sco: the Raman data at T~=~10~K are shown in Fig.\ref{f19}a.
In Fig.~\ref{f19}b we show the ratio of the 2M intensity in $(cc)$ polarization with respect to $(aa)$ configuration as a function of $\omega_{in}$ and in Fig.~\ref{f19}c the resonant Raman excitation profile (RREP) is plotted along with the optical conductivity data provided by the authors of Ref.~\cite{OsafunePRL97}.
For both $(cc)$ and $(aa)$ polarization the resonant enhancement has a maximum around 2.7~eV, about 0.7~eV higher than the CT edge.
The intensity is small for $\omega_{in} < 2$~eV and increases monotonically as the photon energy approaches the CT gap, this increase being followed by a drop for excitations about 3~eV.
The intensity displays an order of magnitude variation as the incident photon energy changes in the visible spectrum.
Besides the correction for the optical response of the spectrometer and detector, by using the complex refractive index derived from ellipsometry and reflectivity measurements, the 'raw' Raman data were also corrected for the optical properties of the material at different wavelengths. % see Chapter 2.

We observe changes in the spectral shape of the 2M as the incident frequency is changed, in the 2LL's case the 2M acquiring sidebands on the high energy side.
These changes are more pronounced in $(aa)$ polarization where for instance the 2.65~eV spectrum (which is close to the edge seen in the $a$-axis conductivity) shows a 2M as a gap-like onset of a continuum.
While the fact that the 2M profile changes substantially with $\omega_{in}$ is also true for 2D cuprates,  one can notice several differences too.
One of them is that the RREP in 2LL's follows more closely the edges of the optical conductivity data.
Moreover, if in the case of cuprates \emph{two} peaks were predicted (and confirmed experimentally) to occur for the 2M peak at $2.8 J$ in the RREP \cite{ChubukovPRL95} (when the incoming energy is in resonance with the bottom and top of the electron-hole continuum) from the data we show in Fig.~\ref{f19}c up to $\omega_{in} = 3.05$ eV we observe only one, rather broad, peak.
It has been argued from numerical diagonalizations of finite clusters \cite{TohyamaPRL02} that this dissimilarity between the 2D cuprates and 2LL's is due to the difference in the spin correlations characterizing the initial and final excited magnetic states, i.e. the weight of the long ranged N\'{e}el type spin-spin correlations in calculating the matrix elements of the current operator plays an important role.

It also turns out that, due to the special topology of 2LL's, a study of the 2M RREP in conjunction with an angular dependence of the 2M intensity in parallel polarization in 2LL's can be helpful for determining a relation between the ratio of the super-exchange integrals $J_{\parallel}$ and $J_{\perp}$ and microscopic parameters like hopping integrals and on site Coulomb interactions \cite{FreitasPRB00}.
Using the effective expression for the photon induced spin exchange coupling mechanism, Eq.~(\ref{e12}), taking into account the anisotropy of the coupling constants denoted by $A$ and $B$ along the rung and leg directions and using the relationship between $H_{FL}$ and the 2D Heisenberg ladder Hamiltonian from Eq.~(\ref{e11}), one can derive the following angular dependence of the 2M intensity for ${\bf e}_{in} \parallel {\bf e}_{out}$: $I_{\parallel} (\omega,\theta) = I (\omega, \theta) [ \cos^{2} (\theta) - \frac{A}{B} \frac{J_{\perp}}{J_{\parallel}} \sin^{2} (\theta) ]$ \cite{FreitasPRB00}.
From this formula, $J_{\perp} / J_{\parallel}$ can be calculated if the $A$ to $B$ ratio is known.
At angles $\theta \neq 0^{\circ}, 90^{\circ}$ from an experimental point of view one has to be careful that the different optical properties of the ladder materials along the $a$ and $c$ axes will induce a non-negligible rotation of the polarization of the incident electric field inside the crystal \cite{GozarPRB02}.
As we see from Fig.~\ref{f19}b the value of $A / B$ is excitation energy dependent and our data suggest that this ratio approaches a constant value in the preresonant regime.
From Fig.~\ref{f19} and using an anisotropy ratio $y = J_{\parallel} / J_{\perp} = 1.25$ (see Fig.~\ref{f17}) we obtain $A / B \approx 2.5$ in the preresonant regime, which would be compatible with an anisotropic local Cu$d$-O$p$ excitation and slightly different hopping parameters along and across the ladder \cite{FreitasPRB00}.

%%%%%%%%%%%%%%%%%%%%%%%%%%%%%%%%%%%%%%%%%%%%%%%%%%%%%%%%%%%
\section{Effects of Temperature and Ca(La) Substitution on the Phononic and Magnetic Excitations in \sco}
%%%%%%%%%%%%%%%%%%%%%%%%%%%%%%%%%%%%%%%%%%%%%%%%%%%%%%%%%%%

%%%%%%%%%%%%%%%%%%%%%%%%%%%%%%%%%%%%%%%%%%%%%%%%%%%%%%%%%%%%
\subsection{Temperature Dependent Electronic and Magnetic Scattering in \sco}
%%%%%%%%%%%%%%%%%%%%%%%%%%%%%%%%%%%%%%%%%%%%%%%%%%%%%%%%%%%%

\begin{figure}[t]
\centerline{
\epsfig{figure=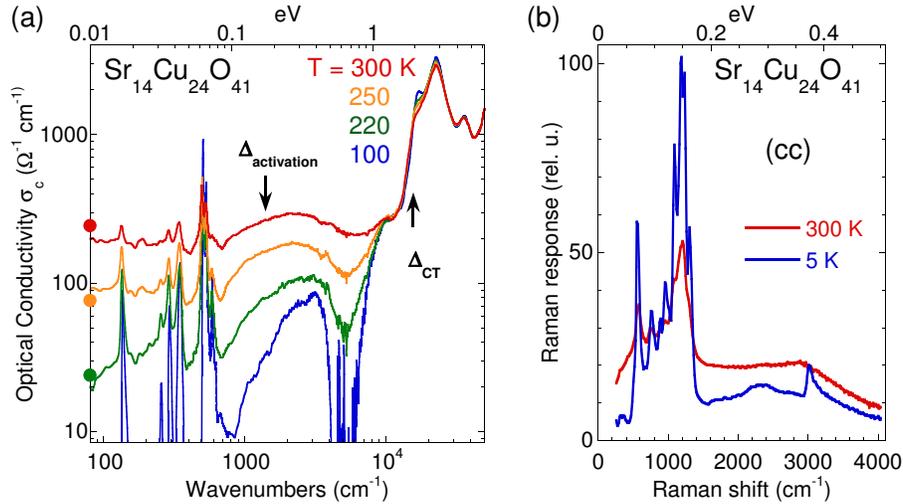,width=120mm}
}
\caption{
(a) Temperature dependence of the $c$-axis optical conductivity (data provided by the authors of Ref.~\cite{EisakiPhysicaC00}) in log-log scale.
The arrows point towards the activation energy determined by $dc$ resistivity above about 150~K and the characteristic energy corresponding to the charge transfer gap.
The circles on the vertical axis represent the $dc$ conductivity values.
(b) Raman response in \sco taken with $\omega_{in} = 1.84$~eV in $(cc)$ polarization at 300 and 5~K.
}
\label{f110}
\end{figure}
The effects of temperature and Ca(La) substitution for Sr discussed in this section set the stage for the following section in which low energy Raman, transport and soft X-ray data argue for the existence of density wave correlations in \slcco compounds.
In Fig.~\ref{f110}a we show the temperature dependence of the $c$-axis conductivity $\sigma_{c} (\omega)$ and in panel (b) the Raman response in \sco for T~=300 and 10~K.
In both IR and Raman data large changes are observed as the \sco crystal is cooled from room temperature.
In Fig.~\ref{f110}a there is a strong suppression of spectral weight below  an energy scale of about 1~eV.
The same figure shows two relevant energy scales of this system: one is the CT gap around 2~eV which was discussed in connection to the resonance properties of the 2M , and the other one is the activation energy inferred from the Arhenius behavior of the $dc$ resistivity above about 150~K \cite{McElfreshPRB89}.
As for the optical sum rule, all the weight is recovered above the CT gap, within an energy scale of $\omega_{c} \approx 3$~eV.
The rapid decrease of the conductivity in the region below 1~eV is correlated to the high activation energy of about 180~meV (= 1450~\cm-1 = 2090~K).
Concomitant to this suppression, which is surprisingly 'uniform' in the 0 to 1~eV range, one observes the development of a broad mid-IR feature and also a sharpening of the phononic features below 1000~\cm-1.
Interestingly, the position of the mid-IR band seems to be close to the semiconducting-like activation energy revealed by the $dc$ resistivity.
Fig.\ref{f110}b shows that a similarly large reduction in the overall intensity of Raman response takes place in an energy range of at least 0.5~eV (4000~\cm-1).
The features which become sharp with cooling are the single and multi-phonon excitations seen around 500, 1200 and 2400~\cm-1 as well as the 2M feature at 3000~\cm-1. 

\begin{figure}[b]
\centerline{
\epsfig{figure=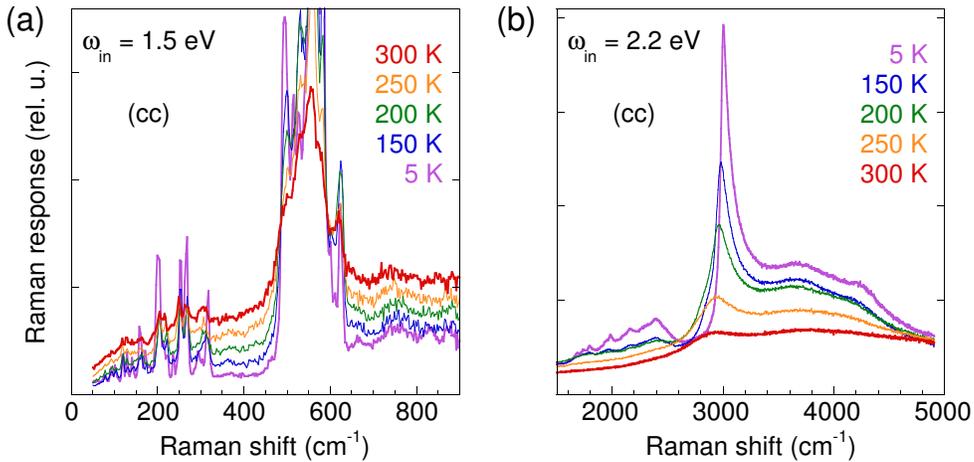,width=130mm}
}
\caption{
Temperature dependent Raman spectra in $(cc)$ polarization for \sco.
(a) Phononic spectra taken with $\omega_{in} = 1.5$~eV (some phonons are truncated).
(b) The 2M peak at 3000~\cm-1 for different temperatures.
The spectra in this panel are taken using $\omega_{in} = 2.2$~eV
}
\label{f111}
\end{figure}
In Fig.~\ref{f111} we show temperature dependent Raman data in two frequency regions: one below 1000~\cm-1 (panel a) and one around 3000~\cm-1 where the 2M feature lies (panel b).
A different spectral shape than in Figs.~\ref{f16} and \ref{f110} is seen due to resonantly enhanced side band structures (see Fig.~\ref{f19}).
The 2M peak is weak and heavily damped at room temperature.
Upon cooling we notice two main features: firstly, the spectral weight increases by almost an order of magnitude, and secondly, the 2M peak sharpens from a width of about 400~\cm-1 at 300~K to 90~\cm-1 FWHM at T~=~10~K.
Because $J / k_{B}T$ remains a large parameter even at room temperature, the magnitude of the observed effects are surprising.
For example, in 2D cuprates the 2M peak remains well defined even above 600~K \cite{KnollPRB90}.
The side bands around 3660 and 4250~\cm-1 observed for the $\omega_{in} = 2.2$~eV also gain spectral weight, proportionally with the sharp 2M feature.
Fig.~\ref{f111}b shows that these sidebands are situated about 650 and $2 \ \times 650$~\cm-1 from the 3000~\cm-1 resonance.
Taking into account that strong phonon scattering characteristic of O modes is found at this frequency, one may argue that these side bands are due to coupled magnon-phonon scattering and bring evidence for spin-lattice interaction in \sco.
These energy considerations favor this scenario compared to one involving multi-magnon scattering because the magnetic continuum starts lower, at $2 \Delta_{S} = 510$~\cm-1.
The latter interpretation remains however a reasonable possibility because in these higher order processes the spectral weight can integrate from a larger part of the Brillouin zone and the boundary of the 2M continuum is dispersive.

The continuum shown in Fig.~\ref{f111}a also gets suppressed with cooling.
Our data confirms the presence of low lying states at high temperatures, observed also in NMR and $c$ axis conductivity, Refs.~\cite{OsafunePRL99,EisakiPhysicaC00} and Fig.~\ref{f110}.
We observe that there is a sharp onset of scattering around 480~\cm-1, close to twice the spin-gap energy.
The 495~\cm-1 mode has been interpreted as evidence for Raman two-magnon scattering \cite{SugaiPSS99}.
However, the temperature dependence of this mode which follows that of the other phonons, the similar suppression with cooling seen not only below this energy but also at higher energies in the 650 to 900~\cm-1 region and the absence of magnetic field effects contradict this proposal.

\begin{figure}[t]
\centerline{
\epsfig{figure=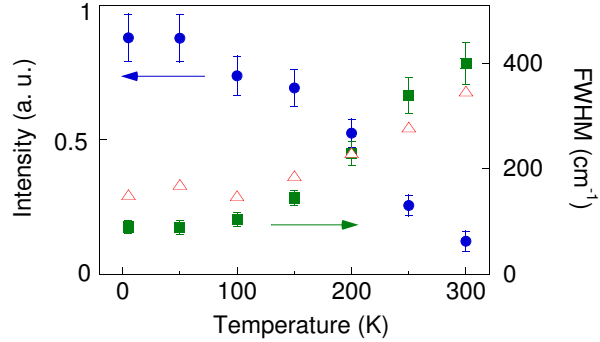,width=80mm}
}
\caption{
The integrated intensity (blue circles, left scale), and the FWHM (green squares, right scale) of the two-magnon peak in \sco from Fig.~\ref{f111}b as a function of temperature.
The triangles represent the continuum intensity (in arbitrary units) around 700~\cm-1 from the data in Fig.~\ref{f111}a.
}
\label{f112}
\end{figure}

The connection between the low and high degrees of freedom in Fig.~\ref{f111}a-b is presented in Fig.~\ref{f112}.
The increase of the electronic Raman background intensity with heating is correlated with the damping of the 2M peak at 3000~\cm-1.
The introduced low energy states reduce the lifetime of the magnetic excitation due to additional relaxational channels provided by the small amount of ladder self-doped carriers.
We note that the drastic changes with temperature take place roughly above 150~K while below this temperature the variation with temperature is much weaker.
This is the temperature at which the $dc$ resistivity changes its activation energy from 2090~K to about half its value, 1345~K \cite{GirshScience02}.
T$^{*}$~=~150~K is also the temperature at which the charge ordering in the chain structures is fully established \cite{FukudaPRB02,RegnaultPRB99} suggesting an interaction between chains and ladders, possibly due to a charge transfer between these systems.
It is possible that this charge transfer takes place also as function of temperature and that it gets suppressed below T$^{*}$.

%%%%%%%%%%%%%%%%%%%%%%%%%%%%%%%%%%%%%%%%%%%%%%%%%%%%
\subsection{The Chain-Ladder Interaction in \sco: Superstructure Effects in the Phononic Spectra}
%%%%%%%%%%%%%%%%%%%%%%%%%%%%%%%%%%%%%%%%%%%%%%%%%%%%

\begin{figure}[t]
\centerline{
\epsfig{figure=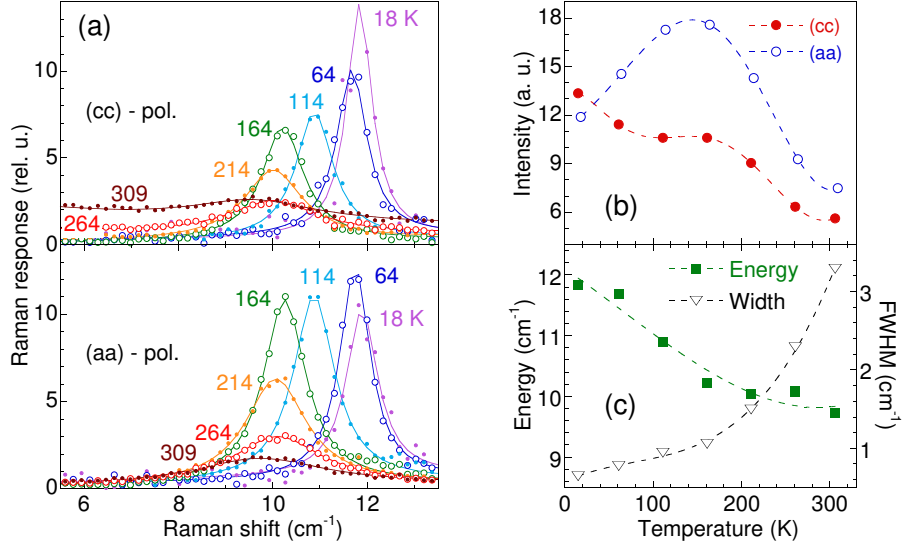,width=120mm}
}
\caption{
(a) Temperature dependence of a low energy \sco phonon taken with $\omega_{in} = 1.65$~eV in $(cc)$ (upper panel) and $(aa)$ (lower panel).
Points represent Raman data and the solid lines are Lorentzian fits.
(b) The dependence on temperature of the phonon intensity in $(cc)$ (filled red circles) and $(aa)$ (empty blue circles) polarizations.
(c) The phonon energy (left scale) and the width (right scale) of the phononic excitation from (a).
Dashed lines in panels (b) and (c) are guides for the eye.
}
\label{f113}
\end{figure}

Raman data in \sco reveals the presence of a very low energy excitation in parallel polarizations.
At low temperatures this mode is found around 12~\cm-1 and we observe a softening of about 20\% with warming up to 300~K.
The temperature dependence of the Raman spectra is shown in Fig.~\ref{f113} for both $(cc)$ and $(aa)$ polarizations.
An excitation at similar energy is seen also in IR absorption data \cite{HomesPrivate} consistent with the lack of inversion symmetry in the \sco crystal.
Applied magnetic fields up to 8~T do not influence the energy of this excitation which suggests that its origin is not magnetic.
This peak is absent in x~=~8 and 12 \scco crystals but it is present around 15~\cm-1 in the \lcco compound \cite{GozarPRL03}.
These properties along with the unusually low energy make us interpret this excitation as a phononic mode associated with the superstructure determined by the chain and the ladder systems.
The chain-ladder commensurability given by the approximate relation $7\ c_{ladder} = 10\ c_{chain}$ will result in a back-folding of the phononic dispersions, which in the case of the acoustic branches will lead to a low energy mode. 
The high effective mass oscillator is understood in this context as a collective motion involving the large number of atoms in the big unit cell of the \sco crystal.

In Fig.~\ref{f113}b-c we plot the temperature dependent energy and width of this low energy phonon.
The crossover below a characteristic temperature of about 120~-~150~K mentioned in the previous subsection is emphasized again by the these data.
The energy of the peak increases rather uniformly with decreasing temperature from 300 to about 15~K but its FWHM shows a variation with temperature which is diminished below 150~K.
The behavior of the integrated intensity of this mode is different in $(cc)$ and $(aa)$ polarizations.
Fig.~\ref{f113}b shows that in $(cc)$ configuration a kink appears about 150~K in the temperature dependent spectral weight while a maximum is seen in the $(aa)$ polarized spectra around this temperature.

In the scenario presented above the presence of the low energy mode Fig.~\ref{f113} is evidence of ladder-chain interaction.
Such an excitation should be sensitive to disorder and even slight modifications in the crystal structure as happens if Sr is substituted by Ca/La.
Symmetry arguments discussed in the next subsection confirm the requirement to consider the full crystal structure for the phononic analysis in \sco and the fact that the disorder introduced by Ca substitution smears out the rich phononic spectra due to the superstructure.
The absence of this mode in Ca substituted crystals thus supports our interpretation.

%%%%%%%%%%%%%%%%%%%%%%%%%%%%%%%%%%%%%%%%%%%%%%%%%%%%
\subsection{Disorder Induced by Ca(La) Substitution}
%%%%%%%%%%%%%%%%%%%%%%%%%%%%%%%%%%%%%%%%%%%%%%%%%%%%

This part deals with the effects of inter Cu-O layers cation substitution.
If Sr is replaced by Ca then the nominal hole concentration in \scco does not change, but what may happen is that the amount of holes in the chain and ladder structures gets redistributed \cite{OsafunePRL97,NuckerPRB00}.
Sr$^{2+}$ substitution by La$^{3+}$ reduces the amount of holes and in \lcco the chains and the ladders are at half filling.
So in analyzing the spin/charge response of the 2LL's one has to consider both the doping and the disorder effects induced by inter-layer cation replacement.

An investigation of these effects is certainly worth pursuing in the context of the constraints imposed by the low dimensionality on the charge dynamics and the occurrence of superconductivity.
Most of the studies in the literature have been focussed on the spin and charge dynamics in pure crystals, although cation substitution is also a source of a random potential.
It is known that in 1D an arbitrary random field localizes all electronic states \cite{AbrikosovAP78} and, in view of the existence of collective excitations of the charge density wave type, % (see section 1.3.4), 
pinning effects due to disorder change qualitatively the $dc$ and the finite frequency transport properties.  
X-ray structural analysis show that the ladder interatomic bonds are modulated upon Sr replacement by Ca \cite{OhtaJPSJ97BougerolPC00} and it was pointed out in a Raman study \cite{OgitaPhysicaB00} that the phononic width increases with Ca concentration in \scco.
Theoretical work shows that the gapped phases of 1D spin systems like 2LL's or dimerized chains are stable against weak disorder and magnetic bond randomness \cite{HymanPRL96OrignacPRB98}.
However, in the doped case, superconductivity in the $d$-channel was found to be destroyed by an arbitrarily small amount of disorder.

\textbf{Ca substitution and phononic scattering --}
\begin{figure}[b]
\centerline{
\epsfig{figure=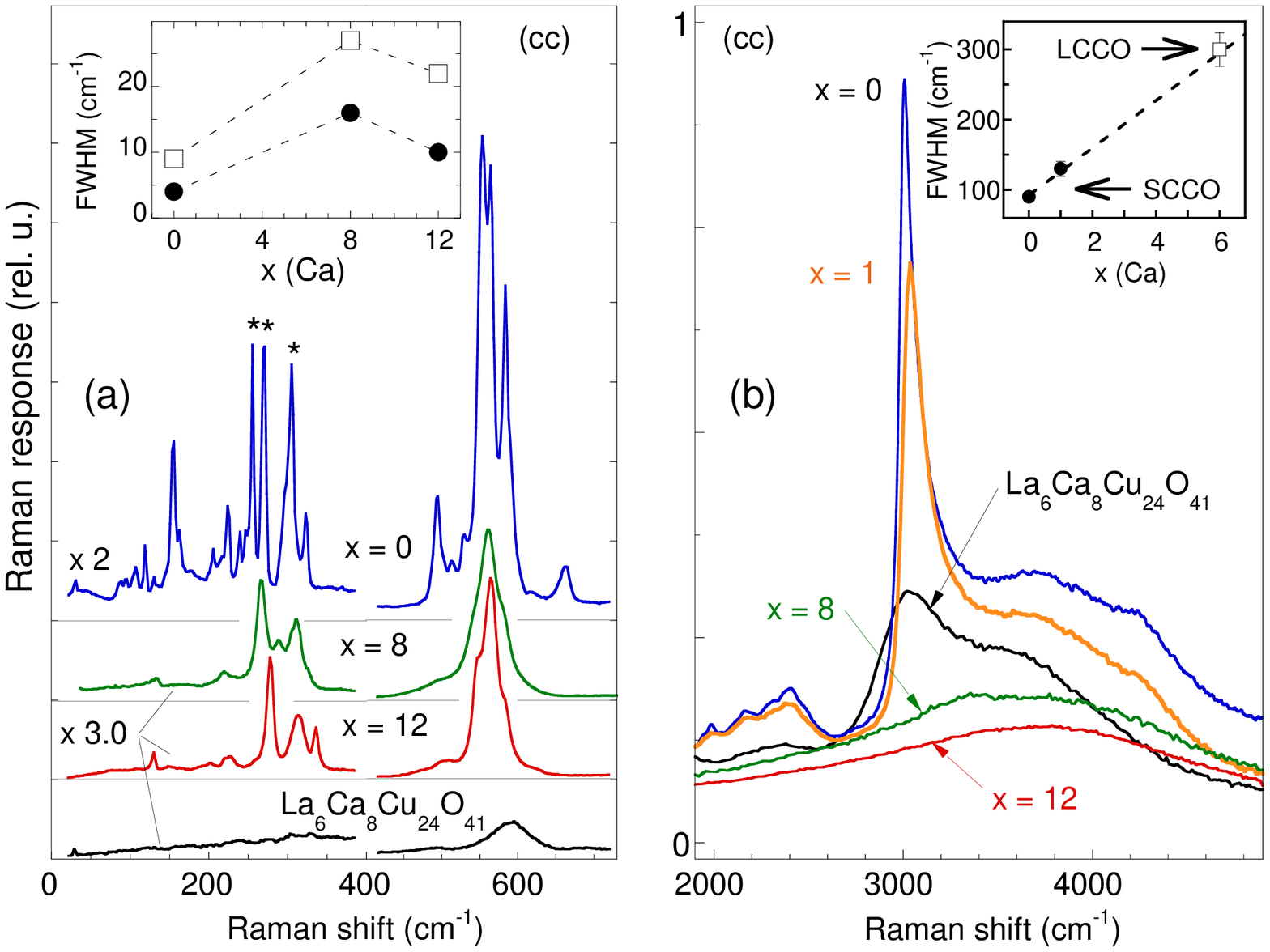,width=120mm}
}
\caption{
(a) T = 10~K Raman data for $\omega_{in} = 2.57$~eV in \lcco and  \scco with x = 0, 8, 12 in $(cc)$ polarization.
The inset shows the full width at half maximum of the 255 and 565~\cm-1 phonons which corresponds to Sr(Ca) (filled circles) and O (empty squares).
Dashed lines are guides for the eye.
(b) Two-magnon spectra taken with $\omega_{in} = 2.2$~eV in $(cc)$ polarization at T~=~10~K in \lcco and \scco with x = 0, 1, 8, 12.
The inset shows the FWHM of the 3000~cm-1 peak as a function of the numbers of Ca atoms in the formula unit.
}
\label{f114}
\end{figure}
If inhomogeneous broadening plays an important role it has to be seen in all the sharp spectroscopic features.
What we try to argue in the following is that the width of both cation and the Cu-O plane modes are renormalized with Ca content.
Fig.~\ref{f114}a shows low temperature phononic Raman spectra in the 0 - 700~\cm-1 energy region.
The data is taken in $(cc)$ polarization with the excitation energy $\omega_{in} = 2.57$~eV; the higher the incoming photon energy the more pronounced is the phononic resonant enhancement.
For \sco we observe a total of 22 clearly resolved phononic modes
extending from 25 to 650~\cm-1.
For \lcco and \scco crystals the features characteristic of O vibrations in the $400 < \omega < 700$~\cm-1 region broaden into an unresolved band and the rich fine structure below $\omega < 400$~\cm-1 is smeared out.
Clear evidence for the interaction between the chain and the ladder structures in \sco can be inferred from symmetry considerations alone.
If these two units were considered separately a total number of six fully symmetric phonons should be observed in $(cc)$ polarization \cite{PopovicPRB00}, three from the chain structure, $Amma$ ($D_{2h}^{17}$) space group, and three from the ladder structure, $Fmmm$ ($D_{2h}^{23}$) space group \cite{McCarronMRB88}.
If one considers the full crystal structure, two 'options' are available.
The first one is to take into account a small displacement of the adjacent Cu-O chains with respect to each other (see Fig.~3 in Ref.~\cite{McCarronMRB88}) and analyze the phonons within the $Pcc2$ ($C_{2v}^{3}$) space group which will give a total of 237 $A_{1}$ modes.
The second one is to neglect this small displacement, as it is the case of Sr$_{8}$Ca$_{6}$Cu$_{24}$O$_{41}$ which belongs to the $Cccm$ ($D_{2h}^{20}$) centered space group \cite{McCarronMRB88} and this approach renders a number of 52 A$_{1g}$ modes.
The 22 observed modes in \sco show that one has to include the chain-ladder interaction and the consideration of the higher $Cccm$ symmetry is sufficient.

Marked with asterisks in Fig.~\ref{f114} are three modes in the region between 250 and 320~\cm-1 which show a blue shift consistent with the lower mass of Ca atoms and the reduction in the lattice constants
upon Ca substitution \cite{KatoPhysicaC96}.
Based on the energy shift and on previous phonon analysis done for the (SrCa)$_{2}$CuO$_{3}$ \cite{YoshidaPRB91} compound we assign the modes to Sr/Ca vibrations.
The FWHM of the 255~\cm-1 phonon in \sco is 4~\cm-1 as compared to 16 and 10~\cm-1 in the x = 8 and 12 \scco samples respectively.
We observe a similar behavior in the phononic modes originating from  Cu-O planes.
Three prominent features are seen in the 550 -- 600~\cm-1 region for the \sco crystal.
We assign the mode with intermediate energy around 565~\cm-1 to O$_{ladder}$ vibration.
The lower and upper modes around 545 and 585~\cm-1 have frequencies close to vibrations of the O atoms in the chains as observed in  (SrCa)$_{2}$CuO$_{3}$ and CuO \cite{PopovicPRB00,YoshidaPRB91} compounds.
Fits for the 550~\cm-1 band in SCCO crystals reveal that the FWHM of the 565~\cm-1 mode increases from 9~\cm-1 for \sco to 27 and 22~\cm-1 for x = 8 and 12 \scco crystals, see the inset of Fig.~\ref{f114}a.
This is similar to what happens to the 255 Ca/Sr mode suggesting that the \scco crystals become again more homogeneous at higher Ca substitution level. 
The data for the LCCO crystal shows that in this material phonons are affected the strongest by disorder which is most likely due to the high La mass and atomic size compared to Ca or Sr atoms.

\textbf{Ca substitution and magnetic scattering --}
\begin{figure}[b]
\centerline{
\epsfig{figure=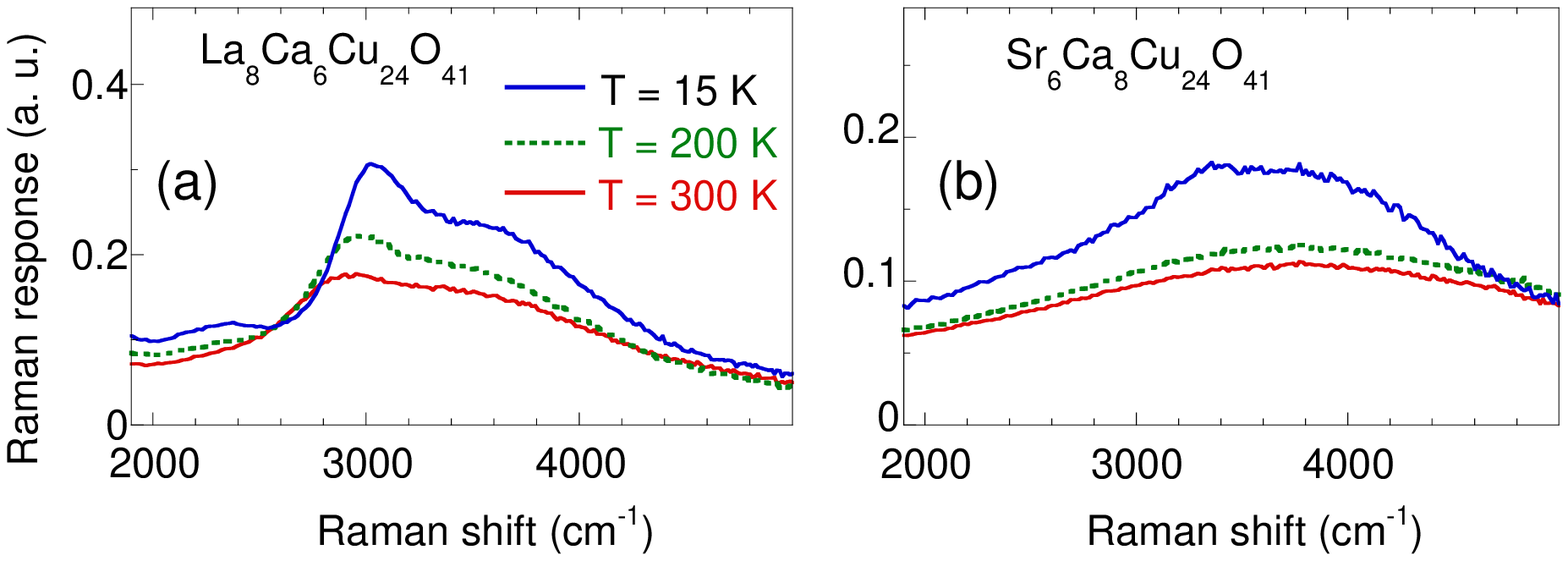,width=130mm}
}
\caption{
Two-magnon scattering in \lcco and x = 8 \scco in $(cc)$ polarization for three temperatures.
}
\label{f115}
\end{figure}
Regarding the sharp 2M Raman resonance, Fig.~\ref{f114}b, one can see dramatic changes taking place with Ca substitution at T~=~10~K and that these changes also affect the 2M sidebands.
In \sco the FWHM is 90~\cm-1.
Ca substitution leads to hardening and to substantial broadening of the magnetic peak accompanied by a drastic decrease in its scattering intensity.
One Ca atom in the formula unit of \scco increases the spectral width by 30\%, see inset of Fig.~\ref{f114}b.
This effect can be ascribed to the intrinsic inhomogeneity rather than a marginal effect on the lattice constants and hole transfer from the chains to the ladders \cite{OsafunePRL97}.
The FWHM in x = 8 \scco and \lcco are about the same within the error bars which is remarkable because the latter is an undoped material so the width of the peak seems not to be related to the presence of carriers in the ladders.
Comparison of our data in \lcco and SrCu$_{2}$O$_{3}$ \cite{GoblingPRB03}, both containing 2LL's at half filling, shows clearly that out-of-plane inhomogeneities have major impact on the magnetic properties of the ladders.

By comparing Fig.~\ref{f111}b and \ref{f114}b
One can also note a resemblance between the effect of temperature in \sco and Ca substitution in \scco.
Fig.~\ref{f115} shows that temperature effects in \lcco and \sccoeight are suppressed compared to \sco.
In this sense one could introduce an 'effective' temperature associated with the cation substitution level.
A comparison to 2D cuprates is again interesting: in the latter case the 2M is broad to start with even in pure materials, but a different number of cation types between the Cu-O layers (higher in insulating Bi$_{2}$Sr$_{2}$Ca$_{0.5}$Y$_{0.5}$Cu$_{2}$O$_{8}$ than for instance La$_{2}$CuO$_{4}$) does not lead to qualitative changes in the 2M width \cite{SugaiPRB90}.

The data in Fig.~\ref{f114}b suggest that an appropriate phenomenological model to describe the ladder Hamiltonian in Ca doped crystals is $H = \sum_{leg} J^{ij}_{||} {\bf S}_i \cdot {\bf S}_j + \sum_{rung} J^{ij}_{\perp} {\bf S}_i \cdot {\bf S}_j$ where the super-exchange integrals $J^{ij}$ in the lowest order have a contribution proportional to the relative local atomic displacements ${\bf u}_{ij}$ according to $J^{ij}({\bf u}) = J_{0} + (\nabla J) {\bf u}_{ij}$.
The effects of thermal fluctuations on the super-exchange integrals $J_{ij}$ can be included in a similar phenomenological approach \cite{NoriPRL95} which could explain the strong resemblance between the Ca substitution and temperature seen in Figs.~\ref{f114}b and \ref{f115}.
We expect the ratio $<J_{\perp}> / <J_{||}>$ to change with Ca content as structural studies show that the Cu-O bonds along the rungs are less affected by Ca substitution than the Cu-O bonds parallel to the ladder legs \cite{OhtaJPSJ97BougerolPC00}.
Also the hardening of the magnetic peak from 3000~\cm-1 in \sco to about 3375~\cm-1 in x = 8 \scco is consistent with the reduction in the lattice constants at higher Ca substitutional level which will lead to a higher super-exchange $J$, a parameter very sensitive to the interatomic distances \cite{CooperPRB90}.

%%%%%%%%%%%%%%%%%%%%%%%%%%%%%%%%%%%%%%%%%%%
\section{Density-Wave Correlations in Doped Two-Leg Ladders}
%%%%%%%%%%%%%%%%%%%%%%%%%%%%%%%%%%%%%%%%%%%

%%%%%%%%%%%%%%%%%%%%%%%%%%%%%%%%%%%%%%%%%%%%%%%%%%%%%%%%%%%%%%%%%%
\subsection{Density Waves: Competing Ground State to Superconductivity}
%%%%%%%%%%%%%%%%%%%%%%%%%%%%%%%%%%%%%%%%%%%%%%%%%%%%%%%%%%%%%%%%%%

So far we have been investigating mainly the magnetic properties of 2LL's around half filling factor and analyzed the effects of temperature and Sr substitution especially in terms of their influence on the high energy 2M scattering around 3000~\cm-1.
We observed that both the temperature and the isovalent cation substitution produce drastic changes in the optical and Raman spectra from far IR up to energies of several eV.
These properties, along with the established metal-insulator transition found around 60\% Ca doping, the occurrence of superconductivity and the similarities with 2D cuprates, nurture the hope that a study of low energy physics in \scco may reveal universal aspects related to the nature of the ground states in low dimensional correlated spin $S = 1/2$ systems.
It is the purpose of this section to bring evidence for the existence of density wave correlations in doped 2LL's at all Ca substitution levels \cite{GozarPRL03}.
Ground states with broken translational symmetry have been discussed in the context of low dimensional systems \cite{SachdevScience00}.
Examples are states which display a long ranged oscillation of the charge and/or spin densities as well as ones which acquire a topological bond order due to the modulations of the inter-atomic coupling constants, for example of the super-exchange integrals.
It has been indeed found that the charge density waves (CDW) and superconductivity are the predominant competing ground states and the balance between them is ultimately determined by the microscopic parameters of the theoretical models \cite{DagottoScience96,DagottoRPP99}.

\begin{figure}[t]
\centerline{
\epsfig{figure=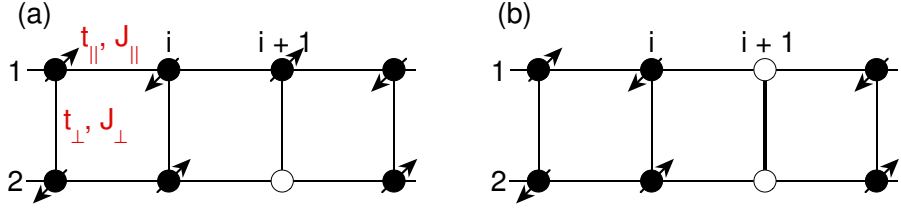,width=120mm}
}
\caption{
Intuitive understanding of the origin of hole pairing in 2LL's in the strong coupling limit ($J_{\perp} \gg J_{\parallel}$).
If an initial hole sits on rung 'i+1' (panel a) the second added hole (panel b) sits on the same rung in order to minimize the number of broken AF bonds.  
}
\label{f116}
\end{figure}

So, what are the low energy excitations one expects from a doped 2LL?
Most of the theoretical studies of 2LL's consist of numerical evaluations, especially exact diagonalization (ED) and density matrix renormalization group techniques (DMRG), performed within the $t_{\parallel} - J_{\parallel}, t_{\perp} - J_{\perp}$ model, see Fig.~\ref{f116}, but not taking in to account the long range Coulomb interactions.
It is interesting to discuss first the cases corresponding to only one or two holes in the ladder structure.
If one hole is present on a ladder rung (Fig.~\ref{f116}a) it can sit on a bonding or antibonding orbital.
Hopping will lead to bands separated roughly by $2 t_{\perp}$ and a bandwidth proportional to $t_{\parallel}$ \cite{TroyerPRB96}.
How tight is the charge bound to the remaining free spin?
This question is connected to the problem of possible spin-charge separation.
Evaluations of hole-spin correlations on a $2 \times 10$ cluster suggest that the unpaired spin remains tightly bound to the injected hole \cite{TroyerPRB96}, so that this composite state carries both charge and spin, in this sense being similar to a quasi-particle.
This is in contrast with the spin-charge separation in the 1D AF chain.

If two holes are present (Fig.~\ref{f116}b) there appears a property which seems to be very robust for 2LL's: pairing.
The following discussion can be intuitively understood starting from the strong coupling limit but studies of finite clusters within the $t_{\parallel} - J_{\parallel}, t_{\perp} - J_{\perp}$ model show that this qualitative discussion holds to the relevant isotropic limit $J = J_{\parallel} = J_{\perp}$ and $t = t_{\parallel} = t_{\perp}$.
If one additional hole is injected in the ladder, it will tend to align on the same ladder rung, see Fig.~\ref{f116}b, in order to minimize the magnetic energy \cite{DagottoScience96,DagottoRPP99}.
The lowest band will be generated by the coherent propagation of hole pairs and it is found in the spin singlet channel.
At finite energies there will be continua of electronic states generated by breaking the pairs, the singlet and the triplet states being almost degenerate when the holes are far apart \cite{TroyerPRB96}.
One can note that in the case of 2LL's it is the purely spin-spin correlations which effectively lead to hole pairing and not an explicit hole-hole attractive interaction and  also that the main energy gain due to pairing is given by the magnitude of the spin gap.
The 'easy' pairing and the kinetic energy gain of the paired holes when pairs are far apart from each other is a non-trivial difference with respect to the 2D cuprates in the sense that in the latter case evaluations prompted by the above arguments lead to macroscopic phase separation.

Since the spin gap $\Delta_{S}$ is to some degree a measure of the hole binding energy it is interesting to discuss what is its evolution with doping.
In the undoped case the lowest triplet excitation is the branch with a minimum at $\pi$ shown in Fig.~\ref{f15} and its magnitude is governed by $J_{\perp}$.
The spin gap remains substantial at isotropic coupling, relevant for experiments, and in this case it is known exactly to be $J / 2$ in the model of Eq.~(\ref{e11}). 
This excitation evolves continuously with doping.
For instance, calculations on a $2 \times 24$ cluster at $1/8$ doping and isotropic coupling shows that the spin gap is about $0.275 J$, about half of the value in the undoped case \cite{PoilblancPRL03}.
Interestingly, pairing generates a different type of singlet-triplet transition \cite{TroyerPRB96,PoilblancPRL03}.
This excitation, present only in the doped case, will consist of breaking of a singlet hole pair into two separate quasi-particles in the triplet channel.
The different kinetic energy gain of the separate holes versus the magnon in the undoped case will lead to different energies of these two types of magnons.
It was argued \cite{TroyerPRB96} that the spin gap evolves discontinuously in 2LL's because it is the 2$^{nd}$ type of magnon which costs less energy.
Later ED and DMRG work \cite{PoilblancPRB00} confirmed this point and showed that in a relevant parameter range the energy of this new type of spin-gap is smaller than the pair breaking continua because a triplet can hybridize with a state formed by two holes (one in bonding and one in antibonding orbitals) forming bound $S = 1$ magnon-hole states.

Once the stability of the hole pair is confirmed to exist in the relevant ranges of the microscopic parameters, it is up to the estimation of residual interactions between the hole pairs and spins to determine what kind of ground state is chosen.
Superconductivity fluctuations were probed within the $t - J$ model by evaluating numerically the pair-pair correlation function, a measure of the stability of the motion of the hole pair in the spin-gapped phase.
This function, which is to be evaluated in the limit of $l \rightarrow \infty$, is defined  as $P(l) = \frac{1}{N} \sum_{i} <\Delta_{i}^{\dag} \Delta_{i + l}>$  where $\Delta_{i}$ is the pair destruction operator at site 'i' given by $\Delta_{i} = \frac{1}{\sqrt{2}} ( c_{i1,\uparrow} c_{i2,\downarrow} - c_{i1,\downarrow} c_{i2,\uparrow})$ (here the 'c' operators are defined within the subspace of no double occupancy).
Early work showed an increase in the pairing tendency as the ratio $J_{\perp} / J_{\parallel}$ was increased \cite{DagottoPRB92RiceEL93}
It has been found for a $2 \times 30$ cluster at $n = 1/8$ doping that SC correlations are dominant and they decay algebraically with $l$ \cite{HaywardPRL95}.
The exponent was found to be smaller than one while density-density correlations were observed to decrease as $l^{-2}$ implying that SC is the dominant phase.
In the same system, by using Green's function techniques, the frequency and wavevector dependence of the superconducting gap \cite{PoilblancPRL03} showed a structure with nodes, much like the $d$-wave pairing symmetry in 2D cuprates.

Pairing does not necessarily mean superconductivity.
Another possibility is that the bound (or single) holes form a spatially ordered pattern, i.e. a CDW ground state.
It has been argued from DMRG calculations that a phase diagram of the isotropic $t - J$ 2LL's, in a relevant range given for instance by $J / t \ < \ 0.4$, will have as generic phase one with gapped spin modes and gapless charge mode \cite{WhitePRB02}.
This 'C1S0' phase \cite{BalentsPRB96} is characterized by $d$-wave like pairing and $4 k_{F}$ CDW correlations, with superconductivity  being the dominant phase \cite{WhitePRB02}.
Note that this $4 k_{F}$ CDW renders a wavelength which is half of the one in conventional Peierls transition.
Phase separation will occur roughly at values $J / t \ > \ 2.5$ \cite{TroyerPRB96,WhitePRB02}.
These numerics also argue that besides these two phases, there are small fully gapped regions (for both spin and charge sectors), to be found generally at commensurate dopings, where a CDW occurs \cite{WhitePRB02}.
The characteristic wavevector of this state is given by $2(k_{Fb} + k_{Fa})$ where $k_{Fb}$/$k_{Fa}$ stand for the Fermi wavevectors of the bonding/antibonding electronic orbitals, discussed in the paragraph related to the charge dynamics in a ladder with one hole.
Interestingly, a finite spin gap is not found to be crucial for the existence of such a CDW so, if the spin gap determines the pairing, the hole crystal can be made either out of single hole or out of hole pairs \cite{WhitePRB02}.

On the experimental side, in \slcco the study of low energy physics is encumbered, compared to 2D cuprates, by the following 'non-intrinsic' facts:
\begin{itemize}
\item The structure is quite complicated due to the presence of the chains and ladders.
We found that these subsystems interact, so one expects that supermodulation will affect carrier dynamics.
\item \sco has a finite hole concentration in the ladder structure to start with.
Ca substitution (and maybe temperature) redistributes the charges between chains and ladders but up to now there is no accurate quantitative determination of this effect. 
On the contrary, there are conflicting views in the literature \cite{OsafunePRL97,NuckerPRB00}. 
\item The effect of O stoichiometry at the crystal surface may be important in accurately determine the carrier concentration; besides, fresh surfaces are not easy to obtain because these materials do not cleave in the $(ac)$ plane.
\end{itemize}

The problem of what happens with the spin gap in the doped ladder is an open issue from an experimental point of view.
On one hand neutron scattering finds $\Delta_{S} = 32$~meV in both \sco \cite{EcclestonPRL98} and x = 11.5 \scco \cite{KatanoPRL99} which says that the spin gap does not change its value.
On the other hand, from the Knight shift (proportional to the uniform susceptibility) and the spin-lattice relaxation data, NMR measurements find a decrease by about 50\% of the ladder spin gap \cite{MagishiPRB98}.
Mayaffre \emph{et al.}, by using the same technique, tried to relate directly the disappearance of the spin gap to the occurrence of superconductivity under pressure \cite{MayaffreScience98}.
Although a finite spin gap is a central issue which underlies the up to date theories predicting that doped ladders are superconducting, it is still not quite clear what the origin of the discrepancy between the INS and NMR data is.

%%%%%%%%%%%%%%%%%%%%%%%%%%%%%%%%%%%%%%%%%%%%%%%%%%%%%%%%%%%%%%%%%%%%
\subsection{Electromagnetic Response of Charge Density Wave Systems}
%%%%%%%%%%%%%%%%%%%%%%%%%%%%%%%%%%%%%%%%%%%%%%%%%%%%%%%%%%%%%%%%%%%%

The purpose of this section is to discuss the main properties of CDW systems and their characteristic excitations.
In the CDW state a gap opens at the Fermi energy and this is observed in $dc$ transport as a metal insulator transition taking place at T$_{c}$.
Due to the change in the lattice constant there also are new phononic modes allowed in the CDW state.
In real systems, which are not strictly 1D, it is possible that not all the Fermi surface gets gapped, so the metallic behavior can continue below T$_{c}$, as is the case of NbSe$_{3}$.
Since the CDW transition involves ionic motions, it can be directly probed by X-rays or neutron scattering \cite{GrunerBook}.

{\bf Excitations out of the CDW state --}
One feature which can be seen in the optical absorption spectra is due to the excitations of electrons across the CDW gap $2 \Delta$.
This belongs to the single particle channel.
Since the Debye energy is much smaller than the Fermi energy the superconducting gaps from BCS theory are typically smaller than  the gap excitations in the CDW state.
For instance, in blue bronze (K$_{0.3}$MoO$_{3}$) which is one of the most studied quasi-1D CDW materials, this energy is found at about $2 \Delta = 125$~meV \cite{DegiorgiPRB91}.
There are also collective excitations out of the condensate and they are related to the space and time variations of the complex order parameter.
Excitations occur due to both phase (phasons) and amplitude (amplitudons) fluctuations.
The interest is to understand the long wavelength limit of these excitations.
As for the amplitude mode, its energy $\omega_{A}$ in the limit $q \rightarrow 0$ is finite.
An oscillation of the gap amplitude $\delta(\Delta)$ will also lead to an oscillation of the ionic positions $\delta(u)$.
The decrease in the condensation energy, $\delta(E_{cond}) = D(\epsilon_{F}) \delta(\Delta^{2}) / 2$ will be equal to the extra kinetic energy associated with ionic displacements, $M N \omega^{2}_{A} (q = 0) \delta(u^{2}) / 2$ where $M, N$ are the ionic mass and number respectively.
As a result one obtains a finite value for $\omega_{A} (q \rightarrow 0)$.

The situation is different for the long wavelength phase mode.
Such motion is a superposition of electronic charge along with ionic oscillations which leads to a high 'effective mass', $m^{*}$.
In the $q \rightarrow 0$ limit involves a translational motion of the undistorted condensate so it will cost no energy.
Its dispersion in the $q \rightarrow 0$ limit is given by $\omega^{2}_{\Phi} (q) = (m / m^{*}) v^{2}_{F} q^{2}$ \cite{LeeSSC74}.
Since phase fluctuations involve dipole fluctuations due to the displacements of the electronic density with respect to the ions the phason is a feature which will be seen in the real part of the optical conductivity data.
The amplitude mode at $q \rightarrow 0$ does not involve such displacements so it is expected to be a Raman active mode.

\begin{figure}[b]
\centerline{
\epsfig{figure=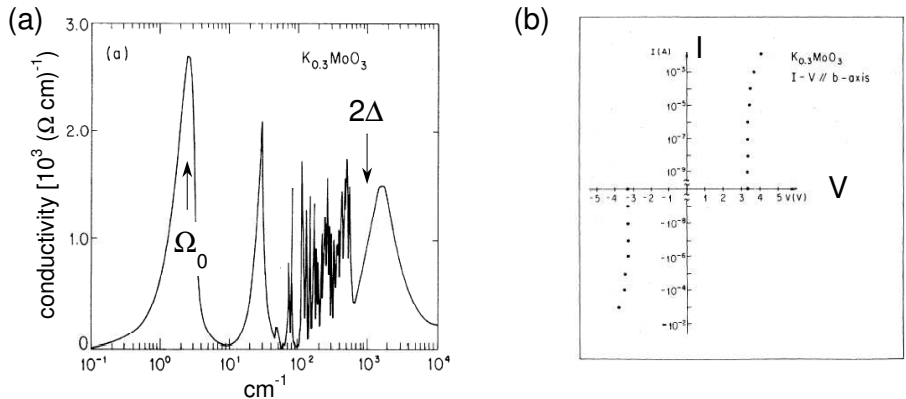,width=120mm}
}
\caption{
Collective excitations and transport in K$_{0.3}$MoO$_{3}$.
(a) Optical conductivity in the CDW phase from Ref.~\cite{DegiorgiPRB91} showing the pinned mode at $\Omega_{0}$ and the single particle edge starting at $2 \Delta$.
Many new electron-phonon coupled modes appear in the mid-IR region below the transition.
(b) The observation of the second threshold, Fr\"{o}hlich superconductivity, in K$_{0.3}$MoO$_{3}$ (from Ref.~\cite{GrunerRMP88}).
}
\label{f119}
\end{figure}

Most interesting is that in the ideal case considered here the phase mode is current carrying and it can slide without friction \cite{Frohlich}.
As a result this excitation will be seen as a $\delta$ function at zero frequency.
The spectral weight of this peak is given by $m / m^{*}$ and this is stolen from the single particle conductivity which becomes an edge, instead of a singularity reflecting the divergence in the density of states \cite{LeeSSC74}, see also Figs. 8 and 9 in Ref.~\cite{DegiorgiPRB91}. 
The interaction with impurities or lattice commensurabilities destroys the infinite conductivity, and the phase mode will be pinned.
As a result, this excitation will be shifted to finite frequencies which characterize the particular impurity potential.
In Fig.~\ref{f119} is shown the example of the blue bronze, the pinning mode as well as the gap feature being seen around 2 and 1000~\cm-1 respectively.

{\bf Zero frequency and microwave transport in the CDW state --}
The existence of a gap and low energy collective excitations leads to several other properties which were seen in $dc$ and finite frequency (typically in the microwave region) conductivity.
In a $I - V$ characteristic one can talk roughly speaking about three regimes.
At low electric fields there is an Ohmic behavior and the conductivity at finite temperatures will be due to thermally excited electrons (normal carriers) out of the condensate.
Above a threshold field, $E_{T}^{(1)}$, related to the magnitude of the pinning potentials, the contribution of the condensate sets in.
The CDW starts moving as a whole and this motion is accomplished through distortions of the phase and/or amplitude of the condensate.
At high fields, above some other threshold field $E_{T}^{(2)}$, the external forces cause a fast sliding motion of the CDW which 'ignores' the underlying pinning potentials and the current increases very steeply (almost infinite differential conductance) for small variations of the applied voltage, see Fig.~\ref{f119}b.
This regime is reminiscent of the ideal case where 'Fr\"{o}hlich superconductivity' should occur. 
The $I - V$ curve in 2$^{nd}$ and 3$^{rd}$ regimes is non-linear and temperature dependent.
Notable is that for an applied $dc$ voltage, the motion of the CDW will also lead in a clean sample to a finite frequency component of the current.
The fundamental frequency of this oscillatory component is directly related to the wavelength of the density wave.

\begin{figure}[t]
\centerline{
\epsfig{figure=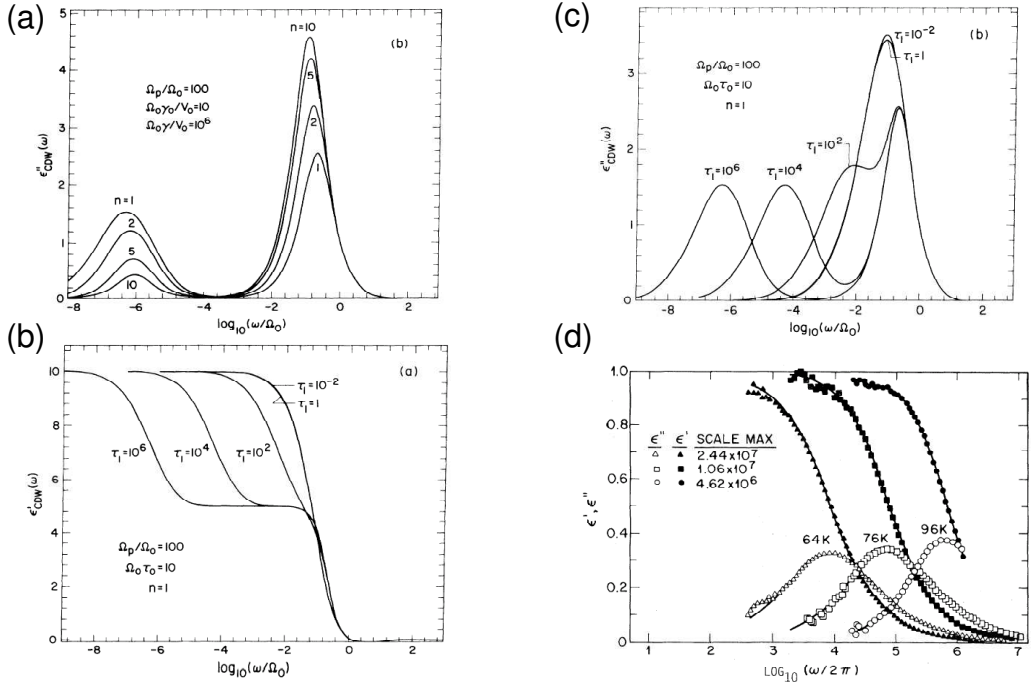,width=140mm}
}
\caption{
The dependence of the real and imaginary part of the CDW dielectric function on frequency on the log scale.
Panels (a), (b) and (c) show theoretical results from Ref.~\cite{LittlewoodPRB87}.
$\Omega_{0}$ and $\Omega_{p}$ are the pinning and the CDW plasma frequencies, $\gamma_{0}$ is an intrinsic damping parameter and $V_{0}$ in panel (a) is the pinning potential.
$\tau_{0}$ and $\tau_{1}$, defined by $\tau_{0} = \gamma_{0} / V_{0}$ and $\tau_{1} = \gamma / V_{0} = \rho^{2}_{c} / \sigma V_{0}$ (with $\sigma$ and $\rho$ being the $dc$ conductivity and the CDW density), are characteristic relaxational times.
The parameter $n$ represents a measure of the distribution in pinning frequencies: $n \rightarrow \infty$ means that there is only one mode in the distribution, the smaller  $n$ is, the broader the distribution.
Panel (d) shows experimental determination of the real and imaginary parts of the dielectric function for three representative temperatures (data from Ref.~\cite{CavaPRB84}).
}
\label{f120}
\end{figure}

{\bf Low frequency CDW relaxation --}
Another low energy feature observed in many well established CDW compounds is a relaxational peak which has a strong temperature dependent energy and damping related to the $dc$ conductivity of the material.
This loss peak is seen typically in the microwave region at energies much lower than the pinning frequency.
For example in K$_{0.3}$MoO$_{3}$ the frequency range is 10$^{4}$ - 10$^{6}$~Hz for temperatures between 50 to 100~K while the pinned mode is roughly at $\Omega_{0} \approx 60$~GHz see Figs.~\ref{f120}d and Figs.~\ref{f119}a respectively.
In Ref.~\cite{LittlewoodPRB87} the author proposes a scenario to reconcile the observations at low and high frequencies, a summary of the results being shown in Fig.~\ref{f120}a-c.
The interpretation of the damped excitation is that it is a longitudinal density wave relaxational mode due to the interaction with normal carriers.
It is argued that this mode, which should not be seen in the transverse channel, is seen however in the dielectric response because of the non-uniform pinning which introduces disorder.
By making the wavevector $k$, according to which the modes can be classified as transverse or  longitudinal, a 'not so good quantum number', disorder mixes the pure longitudinal and transverse character of the excitations.
In other words, breaking of the selection rules make the longitudinal modes appear as poles, rather than zeros, of the dielectric response function.

The main results of the theory in Ref.~\cite{LittlewoodPRB87} are shown in Fig.~\ref{f120} where the CDW dielectric function is plotted as a function of frequency.
The distribution of pinning centers (a measure of disorder) is modeled by a function $g_{n} (x) = (n^{n + 1} / n!) x^{n} \exp(-nx)$ which is peaked at $x = 1$  and satisfies $g_{n \rightarrow \infty} (x) = \delta(x - 1)$.
In Fig.~\ref{f120}a one can see that the disorder leads to the appearance of a mode at lower frequencies which steals spectral weight from the pinning mode situated at the average frequency $\Omega_{0}$.
The stronger the disorder, the higher is the spectral weight redistribution between the two modes.
Panels (b) and (c) in Fig.~\ref{f120} show the real and the imaginary part of the CDW dielectric function for a given $n$.
They are related by Kramers-Kr\"{o}nig relations, so the drop in $Re(\varepsilon)$ leads to a peak in $Im(\varepsilon)$.
These data are plotted for several values of the relaxational time $\tau_{1}$ which mimics (through the dependence on conductivity, see the caption of Fig.~\ref{f120}) a linear variation in temperature.
Decreasing temperature leads to a decrease in conductivity and a higher $\tau_{1}$ and to the softening of the relaxational peak which moves away from $\Omega_{0}$.

{\bf CDW coupling to the uncondensed carriers --} 
Here is a simplified version for the derivation of the longitudinal screening mode shown in Fig.~\ref{f120}.
In this approach the CDW is modeled by an oscillator with a characteristic pinning frequency $\Omega_{0}$ and we neglect internal distortions.
The only other ingredients of the model are the presence of a finite electron density corresponding to thermally activated quasi-particles and the assumption that the interaction between these two fluids is only $via$ an electromagnetic field.
The calculation of the longitudinal CDW modes as well as the coupling to the normal, uncondensed, electrons follows almost identically the treatment of longitudinal phonons and their coupling to plasma oscillations in metals.
In the following, $\vec{u}$ is a uniform displacement of the CDW (in a real crystal this will be within a volume determined by the longitudinal and transverse correlation lengths), $\rho_{c}$ and $m^{*}$ are the CDW charge and mass densities and $\gamma_{0}$ is an intrinsic damping coefficient.
The time derivatives for oscillations at a given frequency $\omega$ are replaced by $\partial / \partial t \rightarrow -i \omega $.
The derivation can be made using the general relations of the Born and Huang model \cite{BornHuang}:
\begin{eqnarray}
- \omega^{2} \vec{u} = - \Omega_{0}^{2} \vec{u} + i \omega \gamma_{0} \vec{u} + \frac{\rho_{c}}{m^{*}} \vec{E}
\label{e14}
\\
\vec{P} = \rho_{c} \vec{u} + \frac{\varepsilon_{\infty} - 1}{4 \pi} \vec{E}
\label{e15}
\end{eqnarray}
Here $\varepsilon_{\infty}$ takes care of the background carrier contributions arising from interband transitions. 
In the absence of carriers, neglecting the damping and using the electrostatic approximation ($\nabla \times \vec{E} = 0$ which means that the field is purely longitudinal and as a result $\vec{E} = \vec{E}_{L}$), these equations allow us to determine the characteristic transverse and longitudinal frequencies.
The equation $- \omega^{2} \vec{u}_{T} = - \Omega_{0}^{2} \vec{u}_{T}$ (because $\vec{E}_{T} = 0$) allows the identification $\Omega_{0} = \Omega_{T}$, i.e. the frequency of the transverse mode.
The longitudinal modes will generate a finite electrostatic field.
Eq.~(\ref{e15}) and Gauss' law $\nabla (\vec{E} + 4 \pi \vec{P}) = 0$ lead to $\nabla (4 \pi \rho_{c} \vec{u}_{L} + \varepsilon_{\infty} \vec{E}) = 0$ so $\vec{E} = - 4 \pi \rho_{c} \vec{u}_{L} / \varepsilon_{\infty}$.
Plugging this relation in Eq.~(\ref{e15}) one obtains $- \omega^{2} \vec{u}_{L} = - \Omega_{0}^{2} \vec{u}_{L} - 4 \pi \rho_{c}^{2} / \varepsilon_{\infty} m^{*} \vec{u}_{L}$ which gives the frequency of the longitudinal mode $\Omega_{in} = \sqrt{\Omega_{0}^{2} + \Omega_{p}^{2} / \varepsilon_{\infty}}$ where the plasma frequency is given by $\Omega_{p}^{2} = 4 \pi \rho_{c}^{2} / m^{*}$.

What is the dynamics of the CDW in an external field $E_{0}$ of frequency $\omega$?
In the transverse channel the force in the right hand side of Eq.~(\ref{e14}) will be $\rho_{c} E_{0} / m^{*}$ leading to $\vec{u}_{T} = [(\rho_{c} / m^{*}) / (- \omega^{2} + \Omega_{0}^{2} - i \omega \gamma_{0})] \vec{E}_{0}$.
Using Eq.~(\ref{e15}), the relation $\varepsilon = 1 + 4 \pi \chi$, where $\chi = P / E$, as well as the fact that the conductivity is given by $\varepsilon (\omega) = 1 + 4 \pi i \sigma / \omega$, one obtains for the collective contribution to the dielectric function and the real part of the conductivity:
\begin{equation}
\varepsilon_{CDW} (\omega) = \frac{\Omega_{p}^{2}}{\Omega_{0}^{2} - i \omega \gamma_{0} - \omega^{2}}
\ \ \ \ \ \ \ \ 
\sigma_{CDW} (\omega) = \frac{1}{4 \pi} \frac{- i \omega \Omega_{p}^{2}}{\Omega_{0}^{2} - i \omega \gamma_{0} - \omega^{2}}
\label{e16}
\end{equation}
These equations will render a peak at the pinning frequency $\Omega_{0}$ in both $\varepsilon (\omega)$ and $\sigma (\omega)$.  

We deal now with the dynamics of the longitudinal modes in the presence of carriers.
One has to worry in this case about the associated internal fields and screening effects. 
One can derive a relation between the CDW displacement $\vec{u}_{L}$ and the local field which should become $\vec{E} = - 4 \pi \rho_{c} \vec{u}_{L} / \varepsilon_{\infty}$ in the limit of zero $dc$ conductivity.
The only difference now is that the first Maxwell equation changes to $\nabla (\vec{E} + 4 \pi \vec{P}) = \rho_{qp}$, where $\rho_{qp}$ is the quasi-particle density.
The continuity equation $- i \omega \rho_{qp} + \nabla \vec{j} = 0$ and Ohm's law $\vec{j} = \sigma_{qp} \vec{E}$ lead to the relation $i \omega \rho = \sigma_{qp} \nabla \vec{E}$ so, using Gauss' law, one obtains $\nabla (4 \pi \sigma_{qp} \vec{E} - i \omega \vec{E} - 4 \pi i \omega \vec{P}) = 0$.
Inserting the expression for polarization from Eq.~(\ref{e15}) and taking into account that we deal with longitudinal fields one obtains:
\begin{equation}
\vec{E} = \frac{4 \pi i \omega \rho_{c}}{4 \pi \sigma_{qp} - i \omega \varepsilon_{\infty}} \vec{u}_{L}
\label{e17}
\end{equation}
Obviously, for $\sigma_{qp} = 0$, Eq.~(\ref{e17}) gives the result of obtained in the previous paragraph in the absence of carriers.
For calculating the longitudinal response, one has thus to replace $\vec{E}$ in (\ref{e14}) with the sum of the external field $\vec{E}_{0}$ and the polarization field given by (\ref{e17}) obtaining a linear relation between $\vec{u}_{L}$ and $\vec{E}_{0}$.
Using (\ref{e15}) one obtains the CDW contribution to the longitudinal dielectric function $\varepsilon_{L}$, which is relevant for Raman scattering, as:
\begin{equation}
\varepsilon_{L} (\omega) = \frac{\Omega_{p}^{2}}{\Omega_{0}^{2} - \omega^{2} - i \gamma_{0} \omega - \frac{i \omega \Omega_{p}^{2}}{4 \pi \sigma_{qp} - i \omega \varepsilon_{\infty}}}
\label{e18}
\end{equation}
In the limit of high frequencies this function has a pole at $\sqrt{\Omega_{0}^{2} + \Omega_{p}^{2} / \varepsilon_{\infty}}$ corresponding to the CDW plasmon and which is the energy of the longitudinal collective mode.
In the limit of low frequencies and neglecting the intrinsic damping $\gamma_{0}$, Eq.~(\ref{e18}) reduces to the following relaxational mode:
\begin{equation}
\varepsilon_{L} (\omega) = \frac{A}{1 - i \omega \tau} \ \ \
\mathrm{with}
\ \ A = \frac{\Omega_{p}^{2}}{\Omega_{0}^{2}} \ \ \
\mathrm{and}
\ \ \ \Gamma = \frac{1}{\tau} = 4 \pi \sigma_{qp} \frac{\Omega_{0}^{2} }{\Omega_{p}^{2}} = 4 \pi \sigma_{qp} \frac{1}{\varepsilon_{0} - \varepsilon_{\infty}}
\label{e19}
\end{equation}
Equations (\ref{e16}) and (\ref{e19}) describe the features seen in Fig.~\ref{f120}.
The proportionality in (\ref{e19}) between $\Gamma$ and the $dc$ conductivity is the result of normal carrier backflow which screens the collective polarization and dissipates energy, suffering lattice momentum relaxation.

%%%%%%%%%%%%%%%%%%%%%%%%%%%%%%%%%%
\subsection{Density Waves in \sco}
%%%%%%%%%%%%%%%%%%%%%%%%%%%%%%%%%%

%%%%%%%%%%%%%%%%%%%%%%%%%%%%%%%%%%%%%%%%%%%%%
\subsubsection{Low energy transport and Raman}
%%%%%%%%%%%%%%%%%%%%%%%%%%%%%%%%%%%%%%%%%%%%%

\begin{figure}[t]
\centerline{
\epsfig{figure=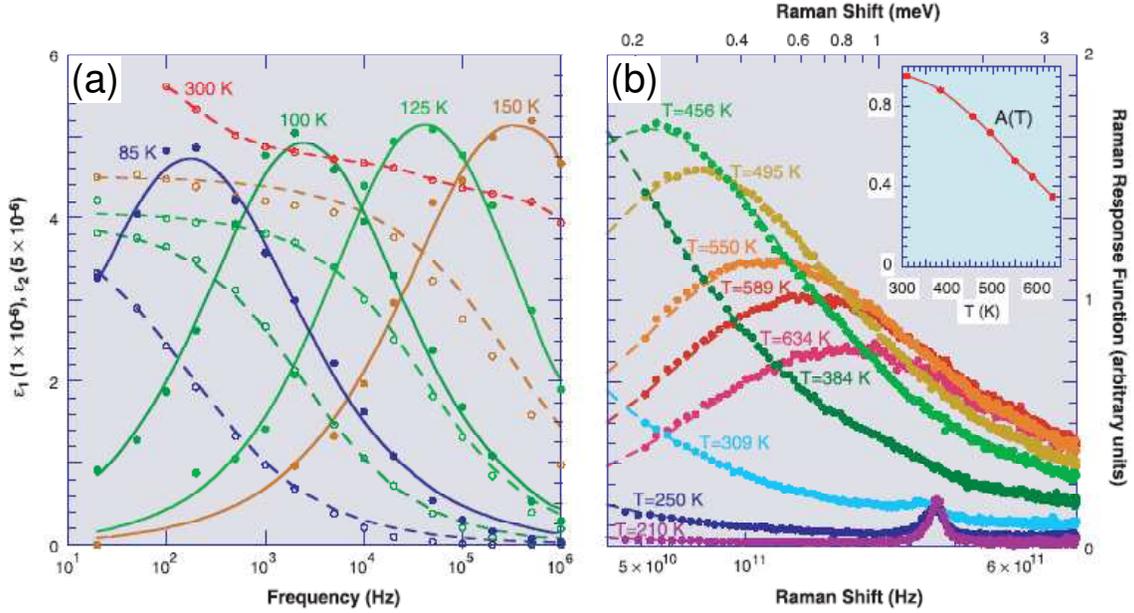,width=150mm}
}
\caption{
(a) The temperature dependence of the real (open circles) and imaginary (solid circles) of the complex dielectric function between 85 and 300~K.
The solid and dashed lines are guides for the eye.
(b) Raman response function in $(cc)$ polarization at several temperatures.
The dots are the data and solid lines are fits to a relaxational behavior as described by Eq.~(\ref{e110}).
The excitation energy used is $\omega_{in} = 1.55$~eV.
The excitation around 350~GHz seen for $T < 300$~K is the phonon shown in Fig.~\ref{f113}. 
The inset shows the temperature dependence of the quasielastic intensity $A(T)$ 
(data from Ref.~\cite{GirshScience02}). 
}
\label{f121}
\end{figure}

In Fig.~\ref{f121}a we show the components of the dielectric response $\varepsilon = \varepsilon_{1} + i \varepsilon_{2}$ as a function of frequency (in log scale) for several temperatures \cite{GirshScience02}.
The imaginary part shows strongly damped, inhomogeneously broadened peaks whose energies are temperature dependent.
These relaxational modes lead to variations in the real part of the dielectric function $\varepsilon_{1}$ up to 300~K and even above.
This data resembles with the dielectric response measured in the CDW compound K$_{0.3}$MoO$_{3}$ which is shown in Fig.~\ref{f120}.
Fig.~\ref{f121}b shows Raman data in a higher temperature range.
Similarly to Fig.~\ref{f121}a we observe an overdamped feature which moves to lower frequencies with cooling.
This excitation disappears below our lower energy cut-off of about 1.5~\cm-1 (equivalent to 50~GHz or 0.185~meV) below about T~=~200~K.
The Raman response function can be well fitted with the expression:
\begin{equation}
\chi'' (\omega, T) = A(T) \frac{\omega \Gamma}{\omega^{2} + \Gamma^{2}} 
\label{e110}
\end{equation}
The temperature dependence of the peak intensity is shown in the inset of Fig.~\ref{f121}.
$A(T)$ decreases by about 60\% from 300 to 640~K.
The temperatures shown in this figure include laser heating effects and they were determined from the ratio of Stokes anti-Stokes spectra for each temperature.

\begin{figure}[b]
\centerline{
\epsfig{figure=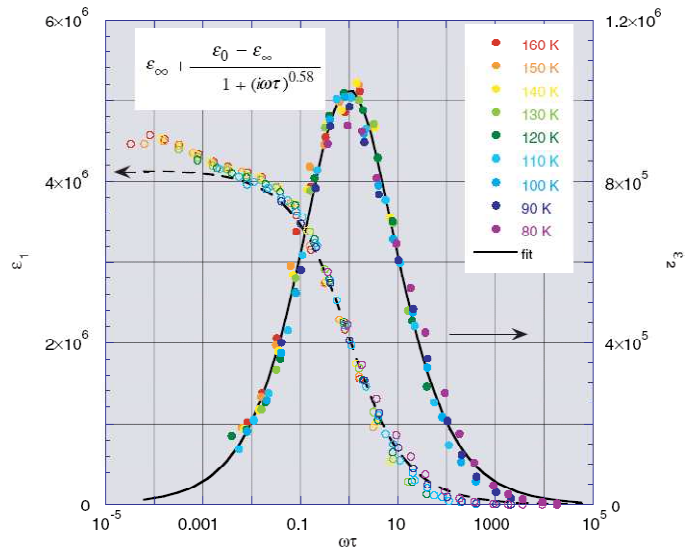,width=90mm}
}
\caption{
Scaling of the complex dielectric function $\varepsilon = \varepsilon_{1} + i \varepsilon_{2}$.
Empty (solid) circles correspond to the real (imaginary) part of $\varepsilon$. (Data from Ref.~\cite{GirshScience02}.) 
}
\label{f122}
\end{figure}

The data in Fig.~\ref{f121}a also allow the extraction of a characteristic transport relaxational time $\tau(T)$ at every temperature by a fit to a relaxational type behavior.
Using this result,  in the entire temperature range the dielectric response between 20~Hz and 10$^{6}$~Hz from Fig.~\ref{f121} can be scaled on a universal generalized Debye relaxational curve given by:
%%%%%%%%%%%%%%%%%%%%%%%
\begin{equation}
\varepsilon(\omega) = \varepsilon_{\infty} + \frac{\varepsilon_{0} - \varepsilon_{\infty}}{1 + [i \omega \tau(T)]^{1 - \alpha}} 
\label{e111}
\end{equation}
The parameter $\alpha$ characterizes the width of the distribution of relaxation times.
The equation for the conventional Debye relaxation has $\alpha = 0$. 
The fit to Eq.~(\ref{e111}) is shown in Fig.~\ref{f122} where the real and imaginary part of $\varepsilon$ is plotted as a function of the dimensionless parameter $\omega \tau$.
The parameter $\alpha$ determined from the fit is $\alpha = 0.42$

\begin{figure}[b]
\centerline{
\epsfig{figure=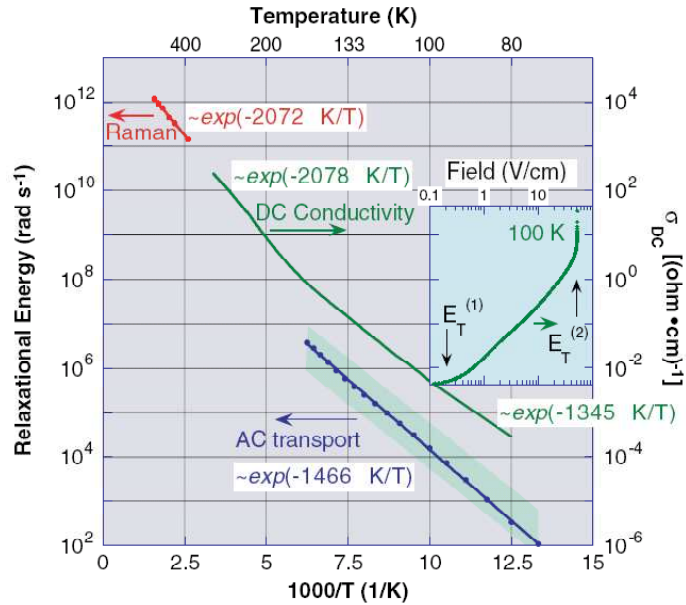,width=90mm}
}
\caption{
Measured $dc$ conductivity (right scale) and the scattering rates (left scales) obtained from fits to the Raman data (using Eq.~(\ref{e110})) and the imaginary parts of the complex dielectric function (using Eq.~(\ref{e111})) shown in Fig.~\ref{f121}. 
Green line: $dc$ conductivity.
Red: the dots are the scattering rates $\Gamma (T)$ from the Raman data in Fig.~\ref{f121}b and the line is an Arhenius fit showing an activated behavior with $\Delta_{Raman} = 2072$~K.
Blue: the dots are the scattering rates $\tau^{-1} (T)$ from the dielectric response in Fig.~\ref{f121}b and the line is a similar exponential fit rendering $\Delta_{\varepsilon} = 1466$~K.
In both regimes the scattering rates show the activated behavior of the $dc$ conductivity.
The shaded area shows the range of scattering rates calculated as described in the text.
The inset shows the nonlinearity in the $dc$ conductivity as a function of applied electric field measured at $T = 100$~K.
Note that the vertical scale for the inset coincides with the vertical scale of the main panel. 
(Data from Ref.~\cite{GirshScience02}.) 
}
\label{f123}
\end{figure}

The temperature dependencies of the relaxational frequencies extracted from the Raman data, $\Gamma(T)$, and from the microwave conductivity data, $\tau^{-1} (T)$, are plotted as a function of inverse temperature in Fig.~\ref{f123}.
On the same plot we show the Arhenius behavior of the $dc$ conductivity.
The $dc$ conductivity in this figure shows activated behavior and the break around $T^{*} = 150$~K points to the existence of two regimes.
At high temperatures the activation energy we obtained is $\Delta_{dc}^{T > T^{*}} = 2078$~K, consistent with previous results~\cite{McElfreshPRB89}.
A value $\Delta_{dc}^{T < T^{*}} = 1345$~K is obtained at low temperatures.
In this figure we observe that the relaxational frequencies have an activated behavior and that the corresponding activation energies match those of the conductivity both above T$^{*}$ (the Raman data) and below T$^{*}$ (the microwave transport data).
This characteristic temperature at which the $dc$ activation changes was discussed also in the end of section 1.3.1 where we noted that it was related to the increase of the electronic Raman continuum, to the variation of the 2M scattering width and also to the temperature dependent intensity of the chain superstructure peaks seen by X-ray scattering. 

The inset in Fig.~\ref{f123} shows the $dc$ conductivity as a function of the applied field.
The arrows mark two threshold fields.
Below $E_{T}^{(1)} \approx 0.2$~V/cm the conductivity obeys Ohm's law and it has the Arhenius temperature dependence shown in the main panel.
For electric fields above $E_{T}^{(1)}$ the $I - V$ characteristics change from linear to approximately quadratic.
At much higher fields, above 50~V/cm, there is a second threshold which marks a very sharp rise of the current.
The differential conductivity in this regime is very high, more than 10$^{5}$~$\Omega^{-1} cm^{-1}$, an estimate limited by contact effects and most likely carried by inhomogeneous filamentary conduction. 

We turn now to the interpretation of the data shown in Figs.~\ref{f121}, \ref{f122} and \ref{f123}.
We remark that the energy range of the relaxational peaks seen in Fig.~\ref{f121} is much lower than the thermal energy or the magnetic and $dc$ activation gaps.
Therefore, this is incompatible with single-particle type excitation and suggest that the low energy charge dynamics is driven by correlated collective behavior.
We identify this strongly temperature dependent feature to be a CDW relaxational mode in the longitudinal channel, screened due to the interaction with thermally excited quasiparticles, as described in the previous section.
We note that electronic Raman scattering can probe directly the longitudinal channel \cite{MilesBook} because the Raman response function, $\chi''(\omega)$, is proportional to Im$[1 / \varepsilon (\omega)]$,  a quantity proportional to $\varepsilon_{L}$ from Eq.~(\ref{e18}).
We can support in what follows this assignment by quantitative comparison with this simple two-fluid model and by the results of the non-linear conductivity measurements as a function of electric field.
Microwave and millimeter wave spectroscopy \cite{KitanoEL01} supports our assignment.
In the end of this chapter, we discuss recent (and direct) evidence for the existence of CDW correlations in \sco provided by X-ray measurements \cite{PeteNature04}.

\begin{figure}[t]
\centerline{
\epsfig{figure=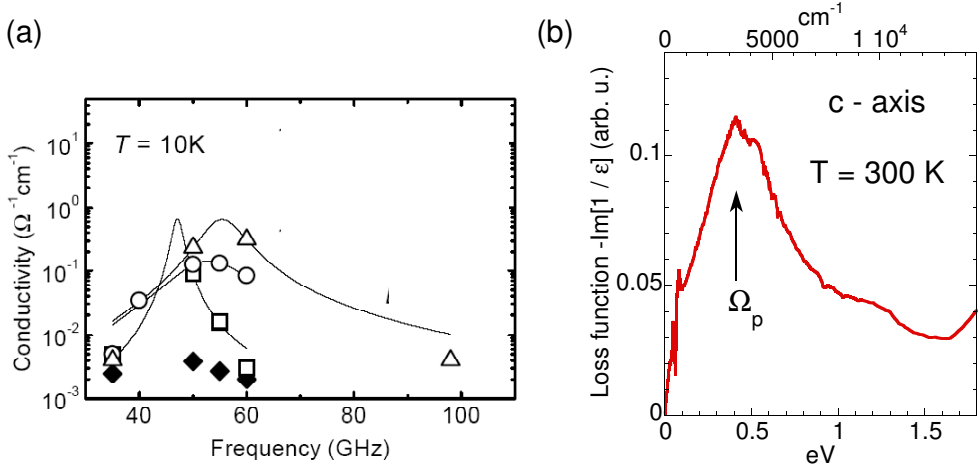,width=130mm}
}
\caption{
(a) Low temperature microwave $c$ ($a$) axis conductivity at T~=~10~K for \sco crystals from three different batches is shown by open (solid) symbols (data from Ref.~\cite{KitanoEL01}).
The solid lines are Lorentzian fits.
(b) The $c$-axis loss function in \sco at 300~K.
$\Omega_{p}$ indicates a plasma edge around 3300~\cm-1. 
}
\label{f124}
\end{figure}

The immediate question prompted by our claim, which essentially ascribes to a common origin our observations in Fig.~\ref{f121} and the properties of K$_{0.3}$MoO$_{3}$ (an established CDW material) shown in Figs.~\ref{f119} and \ref{f120}, is: If we observe a property related to the pinning of an existent CDW, where is the pinned phase mode?
A microwave experiment performed by Kitano \emph{et al.} reported a relatively small and narrow peak between 30 and 70~GHz in the $c$-axis conductivity which was observed up to moderately high temperatures \cite{KitanoEL01}.
The authors attributed this resonance to a collective excitation and speculated about a possible CDW origin.
It turns out that our data along with the results of Kitano \emph{et al.} as well as results of reflectivity measurements form a basis on which these results can be analyzed quantitatively.
In Fig.~\ref{f124} are shown the main result in \cite{KitanoEL01} and the plot of -Im$[1 / \varepsilon (\omega)]$ obtained by our Kramers-Kr\"{o}nig analysis of 'raw' reflectivity data, see Ref.~\cite{OsafunePRL97}.

We believe that the microwave resonance in the 30 to 70~GHz range in Fig.~\ref{f124}a corresponds to the average pinning frequency of the CDW in \sco.
Along with a plasma edge $\Omega_{p} \approx 3300$~\cm-1 extracted from the loss function (see Fig.~\ref{f124}b) and using Eq.~(\ref{e16}) which gives $\varepsilon_{0} - \varepsilon_{\infty} = \Omega_{p}^{2} / \Omega_{0}^{2}$, one obtains for the low frequency dielectric function values of the order of 10$^{6}$, consistent with the experimental observations in Fig.~\ref{f121}.
The two-fluid model described in the previous section, see Eq.~(\ref{e19}), predicts that the relaxational energy is proportional to the activated $dc$ conductivity.
Indeed, the Arhenius behavior of the relaxational energies, extracted both from Raman and transport measurements in Fig.~\ref{f121}, shows from fits with $e^{-\Delta /k_{B} T}$ activation energies similar to those of $dc$ conductivity.
Moreover, we remark that the similarity is not only up to a proportionality factor, but the calculated theoretical values for $\tau^{-1} (T)$ according to Eq.~(\ref{e19}) using the measured values of $\varepsilon_{0}$ and $\sigma_{qp}$ are in agreement with the experiment.
This can be seen in Fig.~\ref{f123} where the calculated values (the shaded area whose thickness takes into account the error bars in the determination of the $dc$ value of the dielectric function $\varepsilon_{1}$) match the measured $\tau^{-1}$ (blue dots). 

The non-linear transport data shown in the inset of Fig.~\ref{f124} for T~=~100~K further confirm the existence of density wave correlations in \sco.
The three regimes observed are typical for systems in which the CDW is pinned by impurities \cite{GrunerBook,GrunerRMP88}.
Below $E_{T}^{(1)}$ the pinned CDW does not contribute to transport and $\sigma_{dc}$ is governed by the quasiparticle response.
Around this value of the field there is an onset of the CDW conductivity due to the relatively slow sliding of the condensate.
In this 2$^{nd}$ regime the predominant damping mechanism is the screening of internal electric fields produced by local CDW deformations by backflow quasi-particle currents.
The 3$^{rd}$ regime defined by fields $E > E_{T}^{(2)}$, indicates a regime of free sliding CDW, the Fr\"{o}hlich superconductivity, also observed in K$_{0.3}$MoO$_{3}$, see Fig.~\ref{f19}.
In this case the velocity of the condensate is so high that it does not feel the background quasi-particle damping.  

The overall consistency among the measured temperature dependencies of the dielectric function, $dc$ conductivity and relaxational energies demonstrates the applicability of the hydrodynamic model description for the low energy carrier dynamics in a CDW ground state.
However, there are several issues which have to be mentioned.
One difference with respect to what happens in well established CDW systems is that the observed relaxational peak in Raman response is at higher energies than the pinned mode at $\Omega_{0}$.
This may be because there is a broad distribution of pinning frequencies and the origin of the Raman relaxational peak is in the high energy side of this distribution.
Up to date there are no measurements of the pinned phase mode at or above 300~K.
Another issue is that although the absolute values of $\tau^{-1} (T)$ calculated according to Eq.~(\ref{e19}) are in agreement with the experiment, the same is not true for the Raman relaxation frequencies $\Gamma (T)$.
The calculated values are about 50 times smaller than the measured ones.
A reduction in the density wave amplitude, as suggested by the decrease in the peak intensity, inset of Fig.~\ref{f121}b, would produce a concomitant increase in $\Gamma$.
Further enhancement in the scattering rate may come from additional relaxational channels due to low lying states which are seen at temperatures higher than about 150~K by magnetic resonance \cite{TakigawaPRB98}, c-axis conductivity (Fig.~\ref{f110}) or Raman scattering (Fig.~\ref{f111}).

The existence of density wave correlations in \sco at temperatures of the order of 650~K gives this compound a distinctive property compared to classical CDW systems.
These high temperatures suggest that in this case it is not the phonons which support the CDW but rather the strong magnetic exchange $J \approx 1300$~K may play an important role in the charge and spin dynamics.
One aspect mentioned in the previous section was that hole pairing in 2LL's is a robust feature due to the AF exchange correlations.
In this respect, an interesting question is: What is the fundamental current carrying object?
Is it due to single or paired electrons?
Helpful in this regard would be to try to measure current oscillations and interference effects (For a description see Chapter 11 in Ref.~\cite{GrunerBook}).
In fact this is probably the only prominent 'classical' transport signature of a CDW state which has not been checked yet in \sco and it would be an interesting project.

%%%%%%%%%%%%%%%%%%%%%%%%%%%%%%%%%%%%%%%%%%%%%
\subsubsection{Soft X-ray scattering from \sco}
%%%%%%%%%%%%%%%%%%%%%%%%%%%%%%%%%%%%%%%%%%%%%

The most direct way to measure CDW ordering is by neutron or X-ray scattering because they can measure directly super-lattice peaks associated with the distortions of the lattice or electronic clouds.
In conventional CDW materials this is the case and the electron-phonon interaction causes atomic displacements and  local electronic density modulations of the order of the atomic numbers.
However, up to date, conventional hard X-ray experiments (using photons with typical energies of the order of tens of keV) failed to detect carrier ordering in the ladder structure of \scco compounds.

Is there any way to observe weak charge modulations which do not involve detectable distortions in the structural lattice?
One way to enhance the scattering amplitude from the doped holes is by exploiting those changes in the optical properties of the materials which occur as a result of doping.
This often involves, as is the case for cuprates, using incident photons with energies about two orders of magnitude smaller than in conventional X-ray experiments.
A real space charge modulation will lead to a proportional change in the Fourier transformed density which in turn is proportional to the dielectric susceptibility of the material, $\chi(k, \omega)$.
The X-ray scattering amplitude is determined by the electronic density and as a result will scale  proportionally to $\chi(k, \omega)$.

It turns out that in 2D cuprates \cite{PeteScience02} and \scco ladders \cite{NuckerPRB00} there are features seen in the X-ray absorption spectra (XAS) which arise directly as a result of hole doping.
The situation is simpler in 2D cuprates and it can be illustrated for La$_{2}$CuO$_{4 + \delta}$: 
For the insulating compounds the oxygen K-edge around 540~eV (which marks the beginning of a continuum of excitations consisting of electron removals from O$1s$ orbitals), has also a prepeak at 538~eV which, due to hybridization, corresponds to intersite O$1s$ $\rightarrow$ Cu$3d$ transitions.
If holes enter O$2p$ orbitals, there will be another prepeak appearing at 535~eV due to the fact that additional O valence states are available to be filled by the excited O1s electron.
The spectral weight of this carrier induced feature is stolen from the 538~eV prepeak.
It is clear that the opening of a new absorption channel at 535~eV will change the optical properties at this energy, in particular of the susceptibility $\chi (k, \omega)$.
This also means that X-ray scattering amplitude for 535~eV incident photons will be enhanced with respect to the non-resonant case by factor proportional to the 'susceptibility contrast' which can be defined as the percentage change of the susceptibility in the doped versus undoped case \cite{PeteScience02}.
Note that this enhancement applies only to the signal from the doped carriers.
\begin{figure}[t]
\centerline{
\epsfig{figure=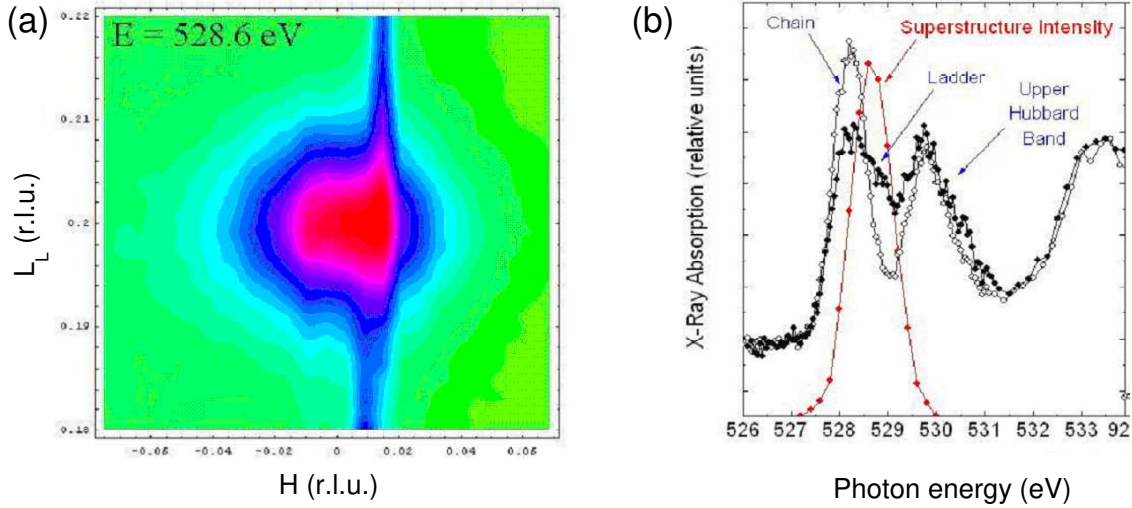,width=150mm}
}
\caption{
(a) The superlattice peak seen at 528.6~eV O$K$ ladder carrier prepeak in the reciprocal space in \sco.
$L_{L}$ on the vertical axis refers is in ladder units.
The rod at $H = 0.01$ is due to the specular reflection from the sample surface.
(b) Black symbols represent X-ray absorption spectra taken with photon polarizations $E \parallel c$ (filled circles) and $E \parallel a$ (empty circles).
The carrier prepeaks corresponding to the chains/ladders are indicated by arrows.
Red symbols are for the integrated intensity of the superlattice reflection seen in panel (a) as a function incident photon frequency.
Data from Ref.~\cite{PeteNature04}.
}
\label{f125}
\end{figure}

In \slcco the XAS spectra have the same general characteristics but the situation is more complicated because the mobile carrier absorption feature is split into chain and ladder features \cite{NuckerPRB00}.
However these excitations can be resolved  and they are shown in Fig.~\ref{f125}b.
This figure shows the characteristic energies of the oxygen K-edge.
The carrier prepeaks are resolved by using different polarizations of the incoming photon fields and one can see that the ladder absorption at 528.6~eV occurs at about 0.5~eV higher energy than the corresponding feature in the chains, consistent with the XAS study in Ref.~\cite{NuckerPRB00}.
A 2D scan in reciprocal space for incident photon energies of 528.6~eV is shown in Fig.~\ref{f125}a.
In this figure the momentum transfer $Q = (2 \pi / a \ H, 2 \pi / b \ K, 2 \pi / c_{L} \ L_{L})$ is in ladder reciprocal units along the $c$-axis.
The vertical line is due to specular reflection from the surface and the displacement from $H = 0$ is due to crystal miscut, the normal to the surface making a finite angle with respect to the $c$-axis.
A superlattice reflection at $(0, 0, 0.2)$ indicates a a charge modulation of 5 ladder units.
In terms of the large crystal structure this momentum transfer corresponds to $L = (c / c_{L}) \ L_{L} = 1.4$, where $c$ and $c_{L}$ are the lattice constants corresponding to the big unit cell and ladder unit cell satisfying $c = 7 c_{L} = 27.3$~\AA \ \cite{McCarronMRB88}.
This Bragg reflection is a true superlattice peak since it does not have the periodicity of the 27.3~\AA \ unit cell and it should not be confused with the five-fold modulation in the chain structures \cite{FukudaPRB02}.

The $(0, 0, 0.2)$ reflection has an unusual excitation profile.
The resonance is shown in Fig.~\ref{f125}b where the energy dependence is plotted along with the absorption spectra.
One can notice that this reflection is seen only in resonance with the ladder absorption at 528.6~eV,  being absent for all other energies, including the oxygen K-edge.
This proves two main aspects: The Bragg peak arises solely from the doped \emph{ladder} holes, and it cannot be due to any structural modulation which would track \emph{all} the features in the O$K$ absorption.
The superlattice peak width in $k$ space gives the correlation lengths $\xi_{c} = 255$~\AA \ and $\xi_{a} = 274$~\AA \ indicating that the order is two dimensional.
This observation is very interesting given the fact that magnetic properties due to the different exchange parameters (Cu-O-Cu bonds making 90$^\circ$ or 180$^\circ$ degrees along the $a$ and $c$ axes respectively, see Fig.~\ref{f11}) as well as the $dc$ transport remain anisotropic, highlighting the importance of inter-ladder Coulomb interactions.

This X-ray scattering study confirms the transport data shown in the previous section in establishing the existence of charge density modulations in doped 2LL's.
The findings are consistent with the predictions of a crystalline order of ladder holes as a competing state to superconductivity \cite{DagottoPRB92RiceEL93,WhitePRB02}.
The absence of structural distortions argues that it is not the conventional electron-phonon interactions, but many-body electronic effects which drive the transition.
One question to address is whether the CDW correlations exist in Ca substituted \sco crystals.
This is the topic of the next section where, based on the similarities with the Raman data in \sco we argue that fluctuations of the density wave order persist at high Ca concentrations and high temperatures.

%%%%%%%%%%%%%%%%%%%%%%%%%%%%%%%%%%%%%%%%%%%%%%%%%%%%%%%%%
\subsection{Signatures of Collective Density Wave Excitations in Doped \scco. Low Energy Raman Data.}
%%%%%%%%%%%%%%%%%%%%%%%%%%%%%%%%%%%%%%%%%%%%%%%%%%%%%%%%%

\begin{figure}[t]
\centerline{
\epsfig{figure=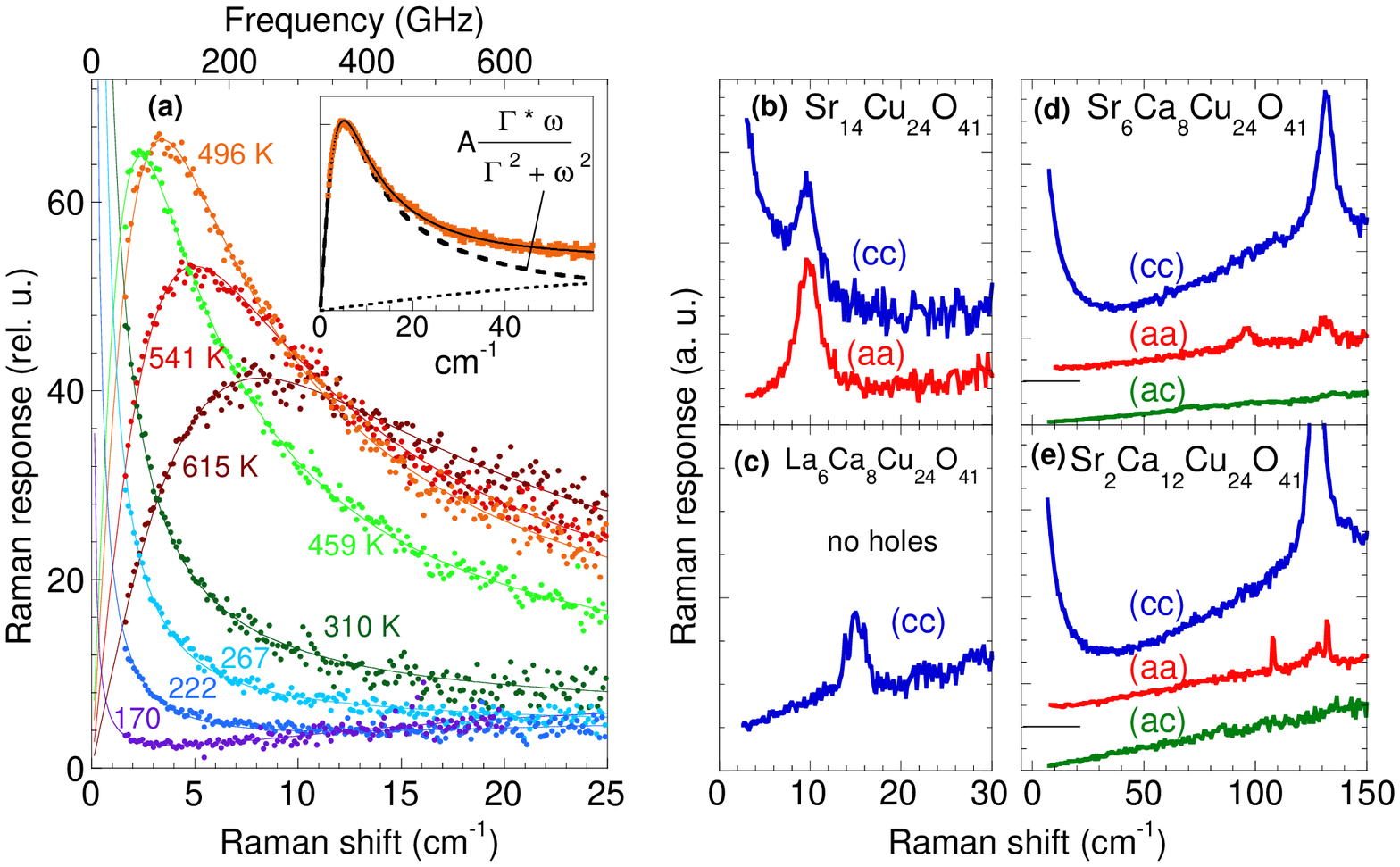,width=130mm}
}
\caption{
A summary of the quasi-elastic Raman scattering in \scco and \lcco compounds.
(a) Temperature dependence of the Raman response in x = 12 \scco in $(cc)$ polarization taken with $\omega_{in} = 1.55$~eV.
This plot is the analogue of Fig.~\ref{f121}b. 
The inset shows a typical fit of the Raman data with a relaxational form, Eq.~(\ref{e110}), and a small contribution from an underlying background.
Panels (b), (c), (d) and (e) show polarized low energy Raman response for \sco, \lcco and x~=~8 and 12 \scco respectively.
Note that the quasi-elastic Raman peak is absent in the undoped \lcco crystal and it is present, only for the polarization parallel to the ladder legs, in all studied \scco samples.
}
\label{f126}
\end{figure}

In Fig.~\ref{f126}a we show low frequency Raman response in x~=~12 \scco at several temperatures.
The $(cc)$ polarized spectra above 300~K are dominated by a quasi-elastic peak, very similar to the one in \sco, see Fig.~\ref{f121}.
The solid lines are fits using the same Eq.~(\ref{e110}) as in Fig.~\ref{f121}.
A small contribution of the background, as shown in the inset, was subtracted. 
The polarization and doping dependence of this relaxational feature are shown in Fig.~\ref{f126}b-e.
We note that the quasi-elastic feature is present only in $(cc)$ polarization and we find it in \scco for all Ca concentrations studied (x~=~0, 8 and 12).
This low energy excitation is absent however in \lcco which contains no holes per formula unit, confirming the fact that it is due to the presence of doped carriers.
We confirmed also that there is no influence of magnetic fields either on this feature or on the modes seen in panels (a) and (c) at 12 and 15~\cm-1 respectively.
This supports the assignment of these modes, shown also in Fig.~\ref{f113} for \sco, to a phonon.

\begin{figure}[t]
\centerline{
\epsfig{figure=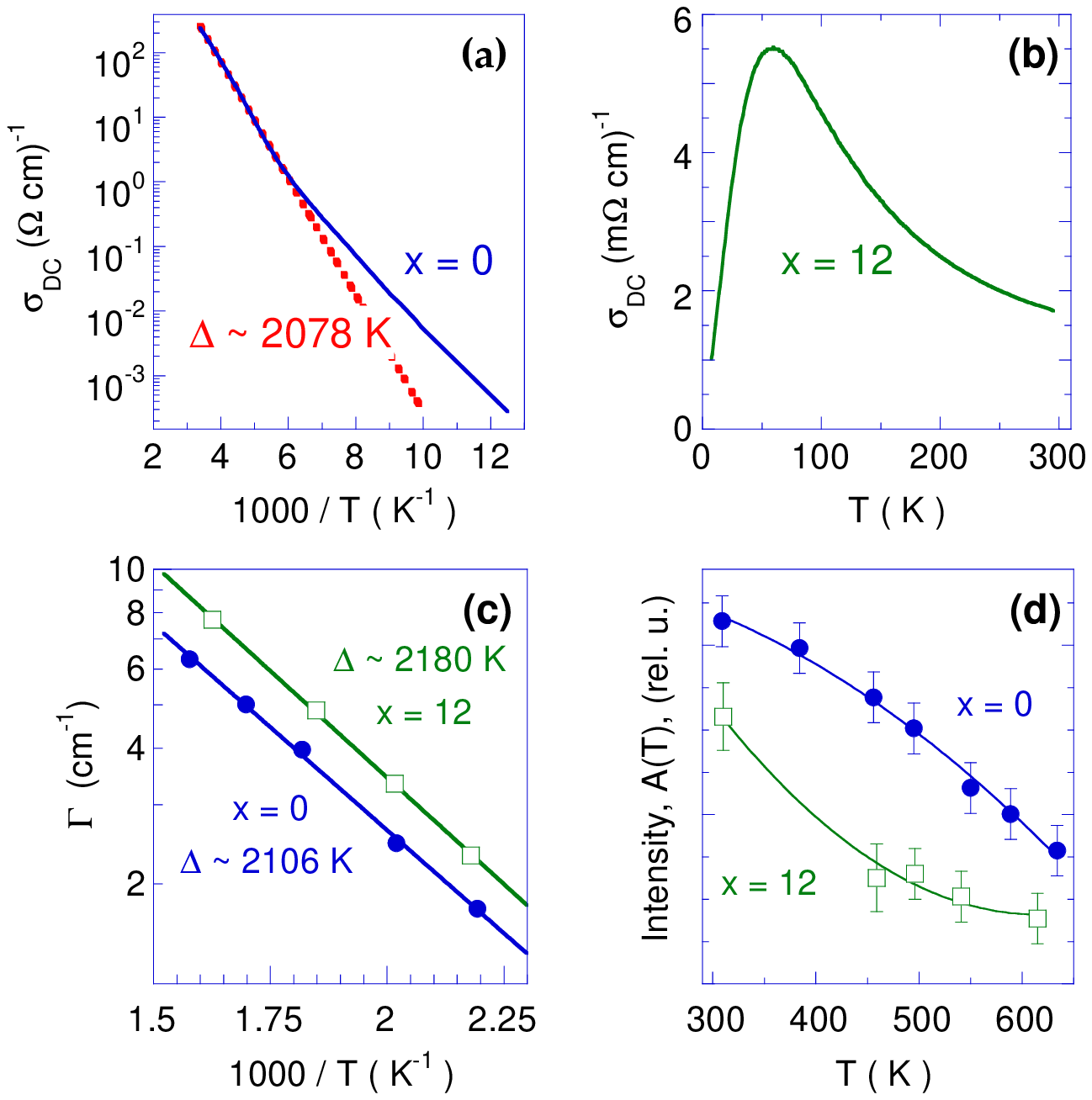,width=90mm}
}
\caption{
Panels (a) and (b) show the temperature dependent $dc$ conductivity for x~=~0 and 12 \scco.
Panel (c) shows Arhenius temperature dependence of the relaxational rate $\Gamma (T)$ for x~=~0 (filled circles) and x~=~12 (empty squares).
The variation of the quasi-elastic peak intensity, $A(T)$, with temperature (solid lines are guides for the eye) is shown in panel (d).
}
\label{f127}
\end{figure}

Interestingly, it turns out that the extracted temperature dependent relaxational energy $\Gamma (T)$ for x~=~12~\scco reveals, similarly to \sco in Fig.~\ref{f123}, an activated behavior of the form $\Gamma (T) \propto \exp(-\Delta /k_{B} T)$.
Moreover, the activation energies are found to be about the same: $\Delta \approx 2100$ and $2070$~K in \sco and x~=~12 \scco, respectively, see Fig.~\ref{f127}c.
While this energy is close to the activation energy of the $dc$ conductivity in \sco, in x~=~12 \scco the temperature dependence of the conductivity is far from exponential, and this can be seen comparing panels (a) and (b) of Fig.~\ref{f127}.
In fact, the behavior shown in panel (b) is very similar to the one in underdoped 2D cuprates: there is a low temperature insulating and a high temperature metallic behavior, in this latter regime the resistivity growing linearly with temperature \cite{OsafunePRL99,BalakirevCM98}.

In the previous paragraphs we argued that the quasi-elastic Raman  scattering in \sco is a signature of collective CDW dynamics.
The main argument in this respect was the Arhenius behavior of the scattering rate with the activation given by the $dc$ transport.  
The low energy scale and the strong similarity between the Raman results in $x = 0$ compared to $x = 8$ and 12 \scco allow us to claim that collective density wave excitations are also present at all Ca substitutional levels.
Confirmation of this scenario comes also from more recent transport and optical conductivity data of Vuleti\'{c} \emph{et al.} \cite{VuleticPRL03} who observe the persistence of the microwave relaxational mode in x~=~3 and 9 \scco.
The authors of this work argue however that Ca substitution suppresses the CDW phase and long range order does not exist above x~=~10.
In this respect we argue that the feature observed in the Raman data in Fig.~\ref{f126} at quite high temperatures in x~=~12 \scco is due to local fluctuations of the CDW order.

How can one reconcile the observation of the same activation energy for $\Gamma (T)$ with the fact that in the insulating regime $\sigma_{dc}$ in x~=~12 \scco is not activated and, moreover, it turns metallic at high temperatures, a behavior clearly not consistent with the prediction of Eq.~(\ref{e19})?
One possible explanation suggested by the $c$-axis optical conductivity data is the following: in \sco one can observe a relatively broad mid-IR peak with an onset around 140~meV, see Fig.~\ref{f110} and Refs.~\cite{OsafunePRL97,EisakiPhysicaC00}.
In \scco this peak continues to be present \cite{OsafunePRL97} and remains a distinct feature although there is a large spectral weight transfer to low energies.
We propose that the common mid-IR feature is responsible for the similarly activated behavior of the relaxation parameter $\Gamma (T)$ and observe that the energy scale of this peak (which is also seen in high T$_{c}$ cuprates) is set by the ladder AF exchange energy of about 135~meV.
In this perspective, a speculative explanation for non-Fermi-liquid like metallic $dc$ conductivity at high Ca substitution levels could be based on a collective density wave contribution.
Ca substitution introduces disorder that could lead to a much broader distribution of pinning frequencies which may extend to very low energies, towards the $dc$ limit, rendering a Fr\"{o}hlich type component contributing to $\sigma_{dc}$.
Intuitively one can imagine that the current carrying objects are not quasi-particles but (because of a small CDW correlation length) 'patches' of holes organized in a density wave order.

Another more conventional scenario for the metallic behavior in x~=~12 \scco could be based on an anisotropic and partially gapped Fermi surface in the context of higher dimensionality of the electronic system.
The soft X-ray study described before, see Fig.~\ref{f125}, shows that the CDW correlations are two dimensional in \sco and recent low frequency dielectric response measurements~\cite{VuleticCM04} were able to track down the relaxational peak in a configuration with the electric field parallel not only to the ladder legs but also to the rung direction.  
One should keep in mind however that the transport along the rung and leg directions is different, as is proven  by the ratio of the $a$ to $c$-axis conductivities, $\rho_{a} / \rho_{c} \approx 10$, for a large range of Ca dopings. 
This can also be related to the fact that we do not observe in Fig.~\ref{f127}a the screened longitudinal CDW relaxational mode in $(aa)$ polarization although the hole ordering is two dimensional. 
Additional support for this conjecture comes from an angle resolved photoemission study \cite{TakahashiPRB97} which shows that while for \sco the gap is finite, for Sr$_{5}$Ca$_9$Cu$_{24}$O$_{41}$ the density of states rises almost to the chemical potential and also from the fact that it is known that the low energy optical spectral weight transfer is enhanced with further increase in Ca substitution \cite{OsafunePRL97}.
In this picture, the insulating behavior in x~=~12 \scco below 70~K can be understood in terms of carrier condensation in the density wave state which leads to a completely gapped Fermi surface.
In order to explain the similar relaxation rates $\Gamma (T)$ for \sco and x~=~12 \scco one has to invoke however a strongly momentum dependent scattering rate and coupling of the condensate to normal carriers.
 
Irrespective of the exact microscopic model, the low energy properties of \scco crystals bring challenging and unresolved aspects.
Moreover, the proof for existence of CDW correlations along with strong similarities between local structural units and transport properties in Cu-O based
ladders and underdoped high-T$_{c}$ materials suggest that carrier dynamics in 2D Cu-O sheets at low hole concentration could be also governed by a collective density wave response.

\section{\bf Summary }

In this chapter we focussed on magnetic and electronic properties of two-leg ladder materials.
We observed at high frequencies (3000~\cm-1) in the \sco compound a two-magnon (2M) resonance characteristic of an undoped ladder which we analyze in terms of symmetry, relaxation and resonance properties.
Our findings regarding the spectral properties of this excitation were contrasted to 2M Raman measurements in other magnetic crystals and existing theoretical calculations, emphasizing the sharpness of the 2M peak in the context of increased quantum fluctuations in one-dimension.
This comparison made us suggest that the spin-spin correlations in an undoped two leg ladder may have a modulated component besides the exponential decay characteristic of a spin liquid ground state. 
We found that the 2M peak resonates with the Mott gap determined by O$2p$ $\rightarrow$ Cu$3d$ transitions, following the behavior of the optical conductivity in the 2-3 eV region.
Interplane Sr substitution for Ca in \sco introduces strong disorder leading to inhomogeneous broadening of the 2M resonance in the undoped system.
The doped holes in the spin liquid ground state further dilute the magnetic correlations, suppressing considerably the spectral weight of this excitation.

\scco crystals at high Ca concentrations are superconducting under pressure and hole pairing was proposed to be a robust feature of doped ladders.
The measured dielectric response in the microwave region, the low energy Raman data, the non-linear transport properties along with soft X-ray scattering allowed us to conclude that the ground state in \scco for a wide range of Ca concentrations ($x \leq 12$) is characterized by charge density wave correlations.
This state seems to be driven not by phonons but by Coulomb forces and many-body effects.
We highlighted the similarity in the finite frequency Raman response as opposed to the very different behavior of the $dc$ resistivity between undoped and doped ladders.
We found that at high Ca concentrations, although the resistivity shows a crossover between insulating and linear in temperature metallic regime, the carrier relaxation is characterized by the same large activation energy ($\approx  2000$~K) which determines the Arhenius behavior of the CDW compound \sco.
This observation prompted us to suggest an unconventional metallic transport driven by collective electronic response.

{\bf Acknowledgments --}
We acknowledge collaboration with P. Abbamonte, B. S. Dennis, M. V. Klein, P. Littlewood, A. Rusydi, and T. Siegrist. 
The ladder crystals were provided by H. Eisaki, N. Motoyama, and S. Uchida.

\end{document}